\documentclass[10pt,aps,prd,twocolumn,notitlepage,nofootinbib,floatfix,superscriptaddress]{revtex4-1}

\usepackage{amsmath, amsthm, amssymb, amsfonts, amsbsy, mathrsfs}

\usepackage{xcolor}
\usepackage{calligra,bm}
\usepackage{stmaryrd}

\usepackage{graphicx}
\graphicspath{{./images/}}

\usepackage[hidelinks,bookmarks=true]{hyperref} 
\hypersetup{pdfstartview=FitH,pdfhighlight=/O,colorlinks=false}

\begin{document}

\title{Dust stars in the minimal exponential measure model}

\author{Reyhan D. Lambaga}\email{d09244004@ntu.edu.tw}
\affiliation{Graduate Institute of Astrophysics, National Taiwan University, Taipei 10617, Taiwan}
\affiliation{Department of Physics and Center for Theoretical Sciences, National Taiwan University, Taipei 10617, Taiwan}
\affiliation{Leung Center for Cosmology and Particle Astrophysics, National Taiwan University, Taipei 10617, Taiwan}
\author{Justin C. Feng}\email{feng@fzu.cz}
\affiliation{Central European Institute for Cosmology and Fundamental 
Physics, Institute of Physics of the Czech Academy of Sciences, Na 
Slovance 1999/2, 182 21 Prague 8, Czech Republic}
\author{Norman Hsia}\email{normanhsia@gmail.com}
\affiliation{Department of Physics and Center for Theoretical Sciences, National Taiwan University, Taipei 10617, Taiwan}
\author{Pisin Chen}
\affiliation{Graduate Institute of Astrophysics, National Taiwan University, Taipei 10617, Taiwan}
\affiliation{Department of Physics and Center for Theoretical Sciences, National Taiwan University, Taipei 10617, Taiwan}
\affiliation{Leung Center for Cosmology and Particle Astrophysics, National Taiwan University, Taipei 10617, Taiwan}
\affiliation{Kavli Institute for Particle Astrophysics and Cosmology, SLAC National Accelerator Laboratory, Stanford University, Stanford, California 94305, USA}



%
%

%
%
\begin{abstract}
We report the existence of horizonless compact object solutions supported by dust in the Minimal Exponential Measure (MEMe) model, a theory which modifies the couplings between gravity and matter without introducing dynamical degrees of freedom. For a perfect fluid source, the field equations for the MEMe model can be rewritten as the Einstein field equations sourced by a perfect fluid with a transformed equation of state, which can endow a sufficiently dense cloud of dust with an effective pressure. The resulting dust-supported horizonless compact objects can have masses below $\sim 10^{-11}~M_\odot$, making them suitable as MACHOs comprising a significant mass fraction for dark matter. A necessary condition for the existence of these compact object solutions is that the single free parameter in the MEMe model is positive-valued. Additionally, we find that this positive sign for the parameter can provide a mechanism for suppressing the formation of (primordial) black holes from the gravitational collapse of matter below a certain mass scale.
\end{abstract}


%
%

\maketitle

%
%

%
%

\section{Introduction}
Despite the successes of general relativity (GR), the large scale behavior of the gravitational field, the fact that gravitational wave observations of compact binary coalescences to date only probe regions up to the photon radius \cite{Cardoso:2017njb,Goebel:1972,Cardoso:2008bp,Berti:2014bla,Glampedakis:2017dvb}, and considerations from the search for an ultraviolet complete theory of quantum gravity all motivate the formulation of extensions to GR \cite{Shankaranarayanan:2022wbx}. The vast majority of extensions involve the introduction of extra dynamical degrees of freedom, but recently, there has been interest in models, called Minimally Modified Gravity (MMG) theories \cite{Lin:2017oow}, that avoid these extra dynamical degrees of freedom. With regard to the counting of dynamical degrees of freedom, one might therefore consider MMGs to be the simplest class of nontrivial modifications to GR. Moreover, MMGs are appealing from a phenomenological perspective, as they can be made to satisfy solar system tests (in which extensions to GR are strongly constrained) without the need for an additional screening mechanism \cite{DeFelice:2020prd,DeFelice:2020eju}.

MMG theories have been classified into two types: Type I MMGs admit an Einstein frame, while Type II MMGs do not \cite{Aoki:2018brq} (this classification has been further refined in \cite{Aoki:2018zcv} according to the behavior of the effective dispersion relations). Much of the work on MMGs have been focused on the Hamiltonian structure \cite{Mukohyama:2019unx} and their large-scale phenomenology \cite{Aoki:2018brq}. While static, spherically symmetric solutions in a Type II MMG were studied in \cite{Aoki:2020oqc}, it is perhaps appropriate to investigate the corresponding solutions in a Type I MMG.

A particularly simple Type I MMG theory is the Minimal Exponential Measure (MEMe) model \cite{Feng:2019dwu}, motivated by semiclassical gravity considerations, in which the matter degrees of freedom couple to an effective metric constructed from the gravitational metric (containing the dynamical tensor degrees of freedom) and an auxiliary rank-2 tensor field, which one may regard as a a field that mediates the interaction between the gravitational metric and the matter fields. The MEMe model, reduces to GR in a vacuum and simplifies considerably for the case of a single species perfect fluid \cite{Feng:2019dwu}; it has been studied in the context of solar system tests \cite{Feng:2020lhp,Danarianto:2023ftf}, the early universe \cite{Feng:2019dwu,Troisi:2025ohh}, at low redshift \cite{Martins:2021zth,Fernandes:2020pxf}, and junction conditions \cite{Feng:2022rga}. Though stellar boundaries \cite{Feng:2022rga} and modifications to white dwarfs \cite{Danarianto:2023rff} have been considered, a detailed study of spherically symmetric solutions in the MEMe model has not been performed.

In this article, we consider perfect fluid solutions of the Einstein frame Tolman-Oppenheimer-Volkoff (TOV) equations for the MEMe model, and find that for a certain sign choice in the single free parameter of the theory, the MEMe model admits solutions describing compact objects supported by dust.
There are no such solutions in GR: For solutions of the TOV equations in GR under a linear (isothermal) equation of state, the energy density scales with $1/r^2$ at large distances \cite{chandrasekhar:1972lre} \cite{chavanis:2002iis}, and the total mass-energy $M$ is therefore infinite. When linear equations of state are considered in the study of stellar structures, the isothermal part of the compact object (e.g. the inner region of a neutron star) is typically contained within an envelope of polytropic matter, so this problem does not arise. 

This paper is organized as follows. We briefly review the MEMe model for a single perfect fluid source in Sec. \ref{sec:meme} and the Tolman-Oppenheimer-Volkoff equation in Sec. \ref{sec:TOV}. We examine in detail the modifications to the equation of state in Sec. \ref{sec:EoS}, and the implications of the resulting energy density limits in Sec. \ref{sec:maxdens}. Numerical solutions to the TOV equation are presented in Sec. \ref{sec:Numerical}, and we consider their stability in Sec. \ref{sec:stabcond}. We discuss the implications of our results in Sec. \ref{sec:disc}.


%
%

%
%
\section{MEMe model}\label{sec:meme}

Here, we review the Minimal Exponential Measure (MEMe) model, which is defined by the action \cite{Feng:2019dwu,Feng:2020lhp,Feng:2022rga}:
\begin{equation}\label{GCA-MEMeAction}
   \begin{aligned}
   S[\varphi,g^{\cdot\cdot},A{_{\cdot}}{^{\cdot}}]= \int d^4x \biggl\{& \frac{1}{2 \kappa}\left[R - 2 \, \tilde{\Lambda} \right]\sqrt{-{g}} \\
   &+ \left(L_{\mathrm{m}}[\varphi,\bar{\mathfrak{g}}^{\cdot\cdot}] - \frac{\lambda}{\kappa} \right) \sqrt{-\mathfrak{g}} \biggr\} ,
   \end{aligned}
\end{equation}
where $\kappa := 8 \pi G$, and the dots indicate index placement. The matter fields $\varphi$ are coupled exclusively to the effective metric $\mathfrak{g}_{\mu \nu}$ and its inverse $\bar{\mathfrak{g}}^{\mu \nu}$, which are in turn constructed from the rank-2 tensor $A{_\mu}{^\alpha}$, the metric $g_{\mu\nu}$, and their respective inverses $\bar{A}{^\alpha}{_\mu}$ and $g^{\mu\nu}$ (indices will be lowered and raised with the metric $g_{\mu\nu}$ and its inverse $g^{\mu\nu}$.):
\begin{equation}\label{GCA-JordanMetric}
\begin{aligned}
   \mathfrak{g}_{\mu \nu} &:= e^{(4-A)/2} \,  A{_\mu}{^\alpha} \, A{_\nu}{^\beta} \, g_{\alpha \beta} ,\\
   \bar{\mathfrak{g}}^{\alpha \beta} &:= e^{-(4-A)/2} \, \bar{A}{^\alpha}{_\mu} \, \bar{A}{^\beta}{_\nu} \, g^{\mu \nu} ,
\end{aligned}
\end{equation}
where $A:=A{_\sigma}{^\sigma}$; later, $|A|$ will be used to denote the determinant of $A{_\mu}{^\alpha}$. One also defines $\tilde{\Lambda}=\Lambda - \lambda$, with $\Lambda$ being the observed value for the cosmological constant.

A few remarks on the physics of this model are in order. Following the reasoning in \cite{Feng:2019dwu}, one can interpret $\lambda$ as a vacuum energy density parameter corresponding to the scale at which the effective geometry described by $\mathfrak{g}_{\mu \nu}$ breaks down. If $\lambda$ is sufficiently large, the MEMe model may require fine-tuning to match cosmological observations,\footnote{The multi-MEMe model described in the appendix may offer a pathway for avoiding such a fine-tuning.} but $\lambda$ need not correspond to the Planck scale, so the degree of fine tuning required may be less extreme than that of the cosmological constant problem in general relativity \cite{Weinberg:1988cp}.

It is convenient to replace the parameter $\lambda$ with the small parameter $q:={\kappa}/{\lambda}$. The variation of Eq. \eqref{GCA-MEMeAction} with respect to ${A}{_\beta}{^\alpha}$ and $g^{\mu\nu}$ yield the respective field equations:
\begin{equation}\label{GCA-ExpFEs}
   \begin{aligned}
   {A}{_\beta}{^\alpha} - \delta{_\beta}{^\alpha} = q \left[ (1/4) \mathfrak{T} \, {A}{_\beta}{^\alpha} - \mathfrak{T}_{\beta \nu} \, \mathfrak{g}^{\alpha \nu} \right],
   \end{aligned}
\end{equation}
\begin{equation}\label{GCA-GEN-GFE-MEMe}
   G_{\mu \nu} +\left[  \Lambda- \, \lambda\left(1 - |A| \right)\right] \, g_{\mu \nu} = \kappa \, |A| \, \bar{A}{^\alpha}{_\mu} \, \bar{A}{^\beta}{_\nu} \, \mathfrak{T}_{\alpha \beta},
\end{equation}
where $G_{\mu \nu}$ is the Einstein tensor and $\mathfrak{T}_{\mu \nu}$ is the energy-momentum tensor obtained from varying $L_\mathrm{m}[\varphi,\bar{\mathfrak{g}}^{\cdot\cdot}]\sqrt{-\mathfrak{g}}$ with respect to $\bar{\mathfrak{g}}^{\mu\nu}$. Equation \eqref{GCA-GEN-GFE-MEMe} has the form of an Einstein equation sourced by a nontrivial energy-momentum tensor; it is in this sense that the MEMe model admits an ``Einstein frame''. Since matter is coupled to 
$\mathfrak{g}_{\mu \nu}$ (and its inverse), one might regard $\mathfrak{g}_{\mu \nu}$ to be a ``Jordan frame'' metric, and $\mathfrak{T}_{\mu \nu}$ to be the ``physical'' energy-momentum tensor.

The field equations simplify when $\mathfrak{T}_{\mu \nu}$ has the form of a perfect fluid:
\begin{equation} \label{GCA-EnergyMomentumPerfectFluid}
   \mathfrak{T}_{\mu \nu} = \left( \hat \rho + \hat p \right)U_\mu U_\nu + \hat p \> \mathfrak{g}_{\mu \nu},
\end{equation}
where $U_\mu$ satisfies $U_\mu U_\nu\bar{\mathfrak{g}}^{\mu\nu}=-1$. For a perfect fluid, one can solve Eq. \eqref{GCA-ExpFEs} exactly to obtain:
\begin{equation}\label{GCA-AnsatzRSYform}
   \begin{aligned}
   A{_\mu}{^\alpha} &= {Y} \delta{_\mu}{^\alpha} - 4(1-Y) u{_\mu} u{^\alpha},\quad
   Y := \frac{4 (1 - \hat p q)}{4 - q \,  (3 \hat p - \hat \rho)}.
   \end{aligned}
\end{equation}
Equation \eqref{GCA-GEN-GFE-MEMe} may then be written in the form:
\begin{equation}\label{EFE}
   G_{\mu \nu} +\Lambda\, g_{\mu \nu} = \kappa T_{\mu \nu},
\end{equation}
where the ``Einstein frame'' energy-momentum tensor $T_{\mu \nu}$ is defined as (with $u^\mu := {U^\mu}/{\sqrt{-U_\sigma U^\sigma}} $):
\begin{equation}\label{GCA-ExpTmnEffDecomp}
   T_{\mu \nu} = \left( \rho + p \right) u_\mu \, u_\nu + p \, g_{\mu \nu} ,
\end{equation}
with $p$ and $\rho$ being the respective ``Einstein frame'' pressures and densities:
\begin{equation}\label{GCA-ExpTmnEffDecompDensityPressure}
   \begin{aligned}
   \rho  = |A| \, (\hat p + \hat \rho) - p, 
   \qquad
   p     = \frac{|A| \, (\hat p \, q - 1) + 1}{q},
   \end{aligned}
\end{equation}
and the determinant $|A|$ having the explicit form:
\begin{equation}\label{GCA-ExpAdetYform}
   |A| = Y^3(4-3Y) = \frac{256 \, (1 - \hat p \, q)^3 (q \, \hat \rho + 1)}{[4 - q \, (3 \hat p - \hat \rho) ]^4}.
\end{equation}
From here on, we shall refer to $\hat \rho$, $\hat p$, $\mathfrak{g}_{\mu \nu}$ and $\mathfrak{T}_{\mu \nu}$ as ``Jordan frame'' quantities, and $\rho$, $p$, ${g}_{\mu \nu}$, and ${T}_{\mu \nu}$ as ``Einstein frame'' quantities.

One can verify that for $\hat\rho, \hat p \ll 1/|q|$, one recovers general relativity coupled to a perfect fluid. If $q<0$, the determinant $|A|$ vanishes when $\hat{\rho}$ approaches a critical density $\hat{\rho}\rightarrow 1/|q|$, and if $q>0$, $|A|\rightarrow 0$ near a critical pressure $\hat{p}\rightarrow 1/q$ (and in the case of a dust, $|A|\rightarrow 0$ for large $\hat{p}$). In this limit, Einstein equations become:
\begin{equation}\label{EFEApprox}
   G_{\mu \nu} \approx -(\Lambda-\kappa/q) g_{\mu \nu}.
\end{equation}
and since $|q|$ is a small parameter, one has a de Sitter or anti de Sitter vacuum with a large curvature parameter.

In \cite{Feng:2019dwu}, the MEMe model was considered for negative\footnote{In the same reference, it was noted that a negative value is preferred to avoid tachyonic instabilities in the case that $A{_\beta}{^\alpha}$ is a fundamentally dynamical degree of freedom, but in this analysis we make no such assumptions about the underlying physics.} $q$, and a recent analysis reveals that a negative-valued $q$ can yield a bounce cosmology \cite{Troisi:2025ohh}. Moreover, since  bosonic and fermionic degrees of freedom respectively contribute positively or negatively to the vacuum energy, one should allow for both positive or negative values for $\lambda=\kappa/q$. For this reason, it is perhaps of interest to consider an extension of the model to accommodate both positive and negative values of $q$. It turns out that one can formulate an extension in a straightforward manner by considering \emph{two} Jordan frame metrics, each associated with a different value of $q$, and splitting the matter sector into two such that each sector is coupled to a different Jordan metric. The details of this extension, termed the multi-MEMe model, are given in the Appendix.


%
%

%
%
\section{TOV equations}\label{sec:TOV}
A general line element for a static, spherically spacetime may be written in the form:
\begin{equation} \label{LineElement}
   ds^{2}= - e^{2\phi} dt^{2} + dr^{2}/\left(1-2m/r\right) + r^2 d\Omega^{2},
\end{equation}
where $d\Omega^{2}=d\theta^2+\sin^2\theta\>d\phi^2$, where $\phi=\phi(r)$ and $m=m(r)$. For a perfect fluid energy-momentum tensor
\begin{equation} \label{EMT}
   T_{\mu\nu}=(\rho+p)u_\mu u_\nu+p g_{\mu\nu},
\end{equation}
where $u^\mu=(e^{-\phi},0,0,0)$, the independent components of the Einstein equations (\ref{EFE}--\ref{GCA-ExpTmnEffDecomp}) reduce to (setting $\Lambda=0$ for simplicity):
\begin{equation} \label{EinsteinSph}
   \begin{aligned}
   m'(r)    &=4\pi \rho(r) r^2,
   \qquad \phi'(r) &=\frac{m(r)+4\pi r^3 p(r)}{[r-2 m(r)]r}.
   \end{aligned}
\end{equation}
The Tolman-Oppenheimer-Volkov (TOV) Equation is:
\begin{equation} \label{TOV0}
   \begin{aligned}
      p'(r) &= -[p(r)+\rho(r)]\phi'(r)
   \end{aligned}
\end{equation}
To close the system equations, one must supply an equation of state (EoS) $p=P(\rho)$ relating $\rho$ and $p$, which is determined by the properties of the fluid in question. More generally, the relation can be supplied by a constraint of the form ${F}({\rho},{p})=0$ (for example $p-P(\rho)=0$). Given some ${F}({\rho},{p})=0$, one may compute the differential $d{F}=0$ to obtain the following equation for $\rho'(r)$:
\begin{equation} \label{rhoeq}
   \begin{aligned}
      \rho'(r) &= -\left(\frac{\partial F}{\partial \rho}\right)^{-1} \left(\frac{\partial F}{\partial p}\right) p'(r) .
   \end{aligned}
\end{equation}
Since Eq. \eqref{rhoeq} is constructed from the differential constraint $d{F}=0$, solutions to Eq. \eqref{rhoeq} will satisfy the constraint ${F}({\rho},{p})=0$ if the initial for $\rho$ and $p$ satisfy ${F}({\rho},{p})=0$. 

\section{Equation of state} \label{sec:EoS}
The field equations of the MEMe model can be described as a perfect fluid with an additional transformation \eqref{GCA-ExpTmnEffDecompDensityPressure} applied to the ``physical'' (Jordan frame) density $\hat\rho$ and pressure $\hat p$. One can solve the TOV equation as one does in general relativity, accounting for the effects of the MEMe model by applying the transformation \eqref{GCA-ExpTmnEffDecompDensityPressure}. If the transformation Eq. \eqref{GCA-ExpTmnEffDecompDensityPressure} can be inverted, then given some EoS $\hat{p}=\hat{P}(\hat\rho)$, one can readily construct an Einstein frame EoS of the form $p=P(\rho)$. However, it is rather difficult to invert Eq. \eqref{GCA-ExpTmnEffDecompDensityPressure} due to factors of the determinant $|A|$. On the other hand, for a simple EoS, one can construct a constraint of the form $F ( \rho,\,p ) = 0$, which we will do explicitly in the following paragraph. In the case where constructing $F(\rho,\,p)=0$ becomes nontrivial, one can use Eq. \eqref{GCA-ExpTmnEffDecompDensityPressure} to perform a change of variables $(\rho,p)\rightarrow(\hat\rho,\hat{p})$; Eq. \eqref{GCA-ExpTmnEffDecompDensityPressure}, and the differential of $\hat{F}(\hat{\rho},\hat{p}):=\hat{p}-\hat{K}(\hat\rho)=0$ form a system of coupled differential equations for $\hat{p}(r)$ and $\hat\rho(r)$; however, such a general method is unnecessary for the simple matter models we consider here.

\begin{figure}
   \includegraphics[width=0.48\textwidth]{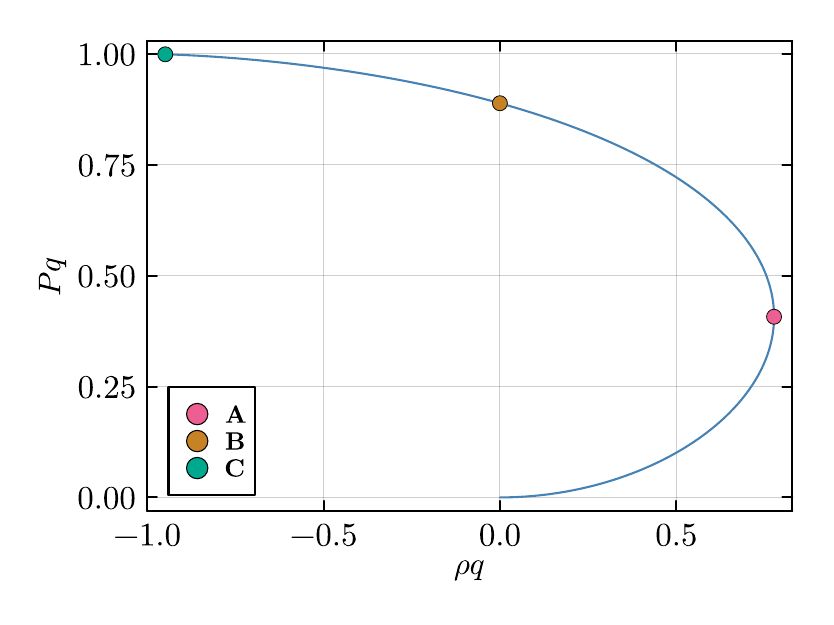}
   \caption{The Einstein frame EoS for a Jordan frame dust, i.e. $\hat{p}=0$, in the case $q>0$. Three regions could be distinguished in the plot: The initial segment up to point $\textbf{A }$,  on which the function is convex; the segment $\textbf{A}\textbf{B}$, on which the density decreases as the pressure increases, but the density is still positive; the segment $\textbf{B}\textbf{C}$, on which the density goes negative. In terms of the Jordan frame density $\hat{\rho}_0$, the point $\textbf{A}$ corresponds to a value $\hat{\rho}_0 = 2/q$, the point $\textbf{B}$ corresponds to $\hat{\rho}_0 = 8/q$, and the point $\textbf{C}$ corresponds to the limit $\hat{\rho}_0\rightarrow+\infty$.}
   \label{fig:EoS}
\end{figure}

We consider a simple EoS of the form:
\begin{equation}\label{linearEOS}
    \hat p=w \hat\rho.
\end{equation}
The corresponding constraint function $F$ may be obtained from Eqs. (\ref{GCA-ExpTmnEffDecompDensityPressure}--\ref{GCA-ExpAdetYform}) by simple algebraic manipulations, as follows. A rearrangement of \eqref{GCA-ExpTmnEffDecompDensityPressure} gives
\begin{equation}\label{GCA-ExpTmnEffDecompDensityPressure2}
   \begin{aligned}
   |A|   & = \frac{\rho+p}{\hat \rho+\hat p}=\frac{p \, q - 1}{\hat p \, q - 1}.
   \end{aligned}
\end{equation}
Substituting \eqref{linearEOS} into the latter equality and rearranging gives the relation 
\begin{equation}\label{rho_in_physical}
    \hat{\rho}(\rho,\,p) = \frac{p + \rho}{1 - q p + w ( 1 + q \rho )},
\end{equation}
which is in turn substituted (together with \eqref{linearEOS}) into \eqref{GCA-ExpAdetYform} and then into either of the equalities in \eqref{GCA-ExpTmnEffDecompDensityPressure} to obtain an implicit relation between $\rho$ and $p$. The result is 
\begin{multline}\label{effEOS}
    F ( \rho,\,p ) = 256 ( 1 +w ) ( 1 - q p )^3 ( 1 + q \rho ) \\
    - ( 4 - 3 q p + q \rho )^4 \big[ 1 - q p + w ( 1 + q \rho ) \big] = 0.
\end{multline}
Now in general, the constraint $F(\rho,\,p)=0$ may yield multiple curves in the $(\rho,p)$ plane. Given $\hat{p}(\hat\rho)$, one can identify the correct curve as the one corresponding to a parameterized curve formed from Eq. \eqref{GCA-ExpTmnEffDecompDensityPressure} and $\hat{p}(\hat\rho)$, with $\hat\rho$ being the parameter. The parameterized curve corresponding to $w=0$ (dust) is plotted in Fig. \ref{fig:EoS}.

The properties of the Einstein frame EoS in the case of negative $q$ have been studied in \cite{Feng:2019dwu}. For $q<0$, we do not find horizonless compact object solutions for a Jordan frame EoS of the form in Eq. \eqref{linearEOS}. In particular, we find that for $q<0$, solutions of the TOV equations admit a mass function $m(r)$ that grows until $2m(r)/r>1$. Intuitively, this follows from the fact that at densities approaching $|1/q|$, the pressure becomes negative; if $p<0$ at $r \sim 0$, then we require that $p'(r)>0$, as one expects a compact object in equilibrium to satisfy $p=0$ at the outer boundary.

We consider now the generic behavior of the Einstein frame EoS in the case of positive $q$. We observe that since the EoS is convex near the origin, the Einstein frame EoS is approximated by a polytropic equation of state when $|P q|\ll1$ of the form $P \approx (3q/8) \rho^2$. Following standard conventions, this corresponds to a polytropic index $n=1$, which is known to yield a stellar radius $R=\pi\sqrt{3 q/2\kappa}$ independent of the density in the Newtonian limit \cite{Chandrasekhar:1939}. On the other hand, the behavior of the Einstein frame EoS becomes rather peculiar for higher pressures. For any physical EoS $\hat{p}=\hat{P}(\hat\rho)$, where $\hat{P}$ is a monotonically increasing function, the Einstein frame density reaches a maximum $\rho_{\rm max}$, which is not greater than $7/9q$. Moreover, the equality $\rho_{\rm max} = 7/9q$ is uniquely attained by (physical) dust. We sketch a proof of these two claims in the next paragraph.

From Eqs. (\ref{GCA-ExpTmnEffDecompDensityPressure}--\ref{GCA-ExpAdetYform}), we obtain 
\begin{equation}\label{rho_explicit}
    \rho(\hat{p},\,\hat{\rho}) = \frac{256 \, (1 - \hat p \, q)^3 (q \, \hat \rho + 1)^2}{q [4 - q \, (3 \hat p - \hat \rho) ]^4} - \frac{1}{q}.
\end{equation}
For $\hat{p} > 1/q$, $\rho(\hat{p},\,\hat{\rho})$ is clearly negative. Otherwise, the inequality of arithmetic and geometric means ($x+y\geq 2\sqrt{xy}$) is applicable to part of the first denominator of Eq. \eqref{rho_explicit}, yielding
\begin{equation}\label{AM-GM}
    4 - q \, (3 \hat p - \hat \rho) \geq 2 \sqrt{3 (1 - \hat p \, q) \cdot (q \, \hat \rho + 1)}
\end{equation}
Taking the fourth power on both sides and substituting into Eq. \eqref{rho_explicit} yields
\begin{equation}\label{rho_bound}
    \rho(\hat{p},\,\hat{\rho}) \leq \frac{7 - 16 \hat p \, q}{9 q} \leq \frac{7}{9 q},
\end{equation}
where the second inequality holds because $\hat{p} = \hat{P}(\hat{\rho})$ is always nonnegative (due to $\hat{P}$ being monotonically increasing.) This bound is saturated at the unique pressure-density point where the inequalities in Eq. \eqref{rho_bound} both take the equal sign. The point is $(\hat{p},\,\hat\rho) = (0,\,2/q)$. The only monotonic EoS that contains this point is $\hat{p} = 0$, the dust EoS. Thus, we conclude that $\rho = 7/9q$ is an upper bound for the Einstein frame density under any monotonically increasing physical EoS, and that this bound is uniquely saturated by Jordan frame dust.


%
%
%
%
%

\section{Maximum density and Black Hole Formation}\label{sec:maxdens}

\begin{figure}
   \includegraphics[width=0.48\textwidth]{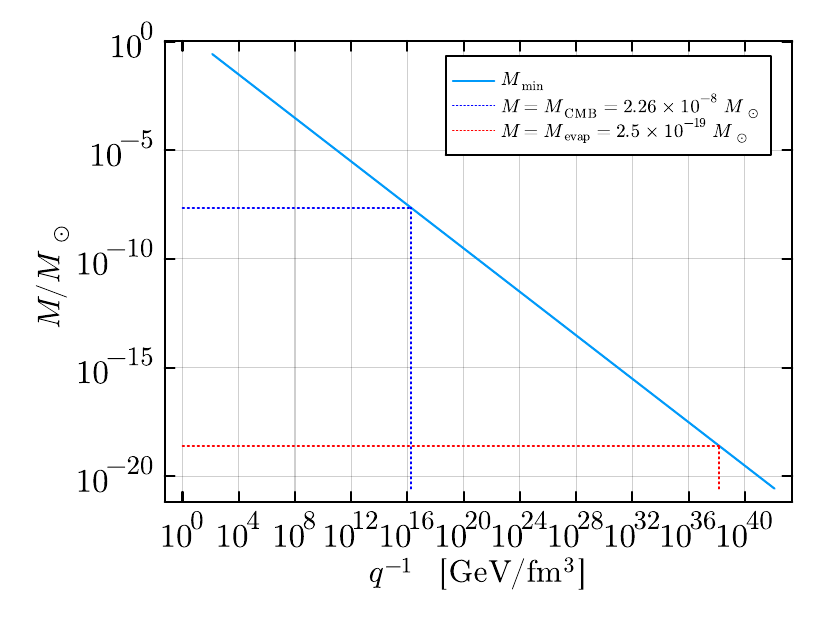}
   \caption{
   A plot of $M_\mathrm{min}$ as a function of $1/q$ (assumed to be positive), given by Eq. \eqref{minmass_in_q}. Several physically significant energy scales and the corresponding values of $M_\mathrm{min}$ are marked by the dotted lines. }
   \label{fig:MinMass}
\end{figure}

An interesting consequence of an upper bound for the Einstein frame energy density is that it
places a rough bound on the minimum mass for a black hole that can be 
formed from gravitational collapse. To see this, consider a spherically
symmetric concentration of mass of uniform density---in general 
relativity, the solution has been extensively studied, and it is 
well-known that the mass $M$, radius $R$, and density $\rho_0$ are related by $\rho_0=3M/4\pi R^3$. It is also well-known from solutions of the TOV equation that in the Buchdahl limit $R \rightarrow R_{\rm Buch} := 9M/4$, the central pressure diverges, and one concludes that the object must collapse to form a black hole. One may then define a critical density $\rho_c$ for a given mass $M$:
\begin{equation} \label{critdens}
   \begin{aligned}
      \rho_{\rm c} := \frac{M}{4\pi R_{\rm Buch}^3} = \frac{16}{243 \pi  M^2},
   \end{aligned}
\end{equation}
which provides a threshold for the collapse; isolated compact objects with overall densities exceeding $\rho_{\rm c}$ at some point during their evolution may be expected to collapse to form a black hole. However, if the Einstein frame density has a maximal value $\rho_{\rm max}$, then $\rho_{\rm c} \leq \rho_{\rm max}$, and it follows that:
\begin{equation} \label{minmass}
   \begin{aligned}
      M \geq M_{\rm min} = \frac{4}{9 \sqrt{3 \pi \rho_{\rm max}}},
   \end{aligned}
\end{equation}
If we take $q$ to be positive, the maximal Einstein frame density attainable under the assumption that the EoS is monotonically increasing was found to be $\rho_{\rm max} = 7/9q$ in Sec. IV. The minimum black hole mass allowed by our theory in this case is therefore (with constants explicitly included):
\begin{equation} \label{minmass_in_q}
   \begin{aligned}
      M_{\rm min} = \frac{c^4}{G^{3/2}} \frac{4}{9} \sqrt{\frac{3 q}{7 \pi}},
   \end{aligned}
\end{equation}
The relationship between $M_{\rm min}$ and energy density scales $1/|q|$ is illustrated in Fig. \ref{fig:MinMass}. 

As discussed in \cite{Feng:2020lhp}, one can obtain a lower bound for the energy density scale $1/|q|$ from the highest energy densities achieved in heavy-ion colliders, which are on the order of $10~\textrm{GeV}/\textrm{fm}^3$ \cite{Pasechnik:2016wkt}.\footnote{In MKS units, $1~\textrm{GeV}/\textrm{fm}^3=1.6\times 10^{35}~\textrm{J}/\textrm{m}^3$, which corresponds to a mass density of $1.8 \times 10^{18}~\mathrm{kg}/\mathrm{m}^3$, about an order of magnitude higher than average density of nuclear matter.} Such a bound corresponds to a value of $M_{\rm min}$ on the order of a solar mass. The absence of anomalous behavior in the observed decays of the heaviest standard model particles may provide a stronger constraint; if one considers the mass scale of the Higgs boson or top quark combined with their Compton wavelengths, one obtains densities on the order of $7\times 10^{11}~\textrm{GeV}/\textrm{fm}^3$, and a lower mass bound of $M_{\mathrm{min}}\sim 10^{-5} M_\odot$. A more detailed study of the effect of the MEMe model on particle interactions may be needed to more precisely establish the lower limit on the energy density scale (which as we see here varies by $\sim 10$ orders of magnitude depending on our assumptions).

\begin{figure}
   \includegraphics[width=0.48\textwidth]{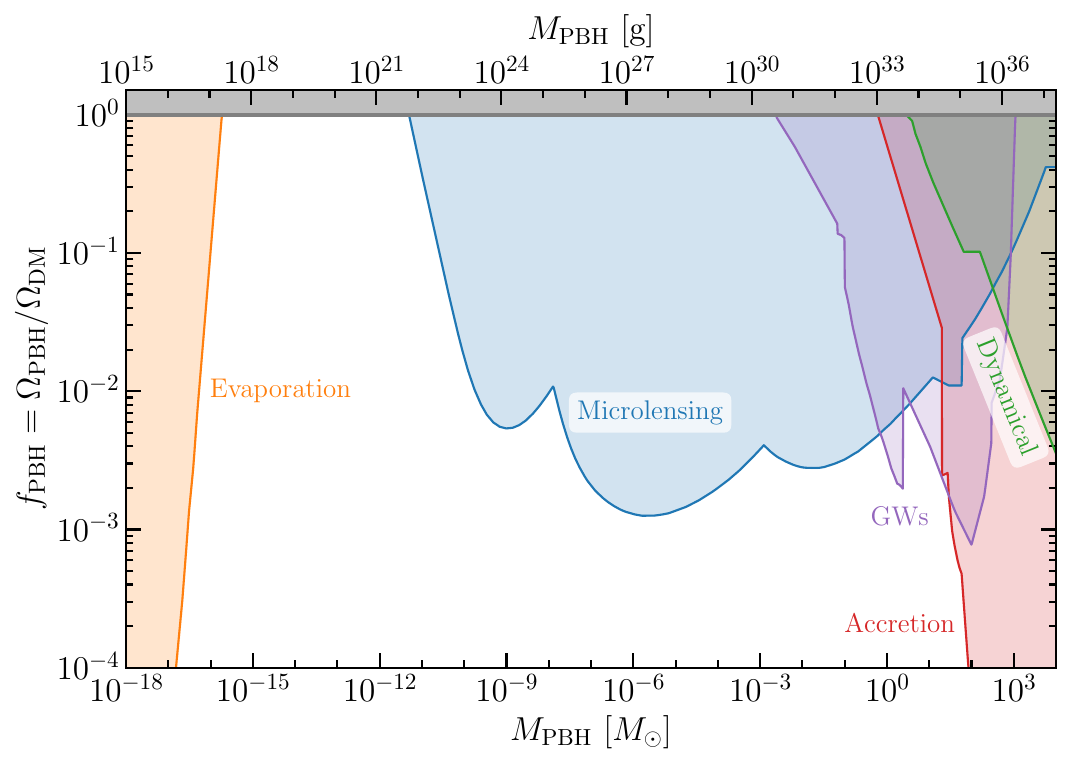}
   \caption{Summary of constraints on the primordial black hole mass fraction of dark matter, some of which (Microlensing, for instance) which also apply to MACHOs. Plot generated using PlotPBHbounds \cite{Kavanagh:2019}, using data current as of May 12, 2025.}
   \label{fig:PBHConstraints}
\end{figure}

If one postulates that the absence of observed signals from the late stages of Primordial Black Hole (PBH) evaporation is a consequence of our lower bound on black hole mass \eqref{minmass_in_q}, one can also obtain an upper bound for the energy density scale $1/|q|$. From \cite{Carr:2020gox}, evaporation constraints place a hard lower bound of $\sim 10^{-19}~M_\odot$ from $M_\mathrm{min}$, corresponding to a upper bound for the energy density $1/|q| \sim 10^{39}~\mathrm{GeV}/\mathrm{fm}^3$ (Fig. \ref{fig:PBHConstraints} suggests a tighter limit of $\sim 10^{-17}~M_\odot$, corresponding to a slightly reduced bound of $1/|q| \sim 10^{35}~\mathrm{GeV}/\mathrm{fm}^3$). A stronger constraint comes from assuming that PBHs contribute significantly to the dark matter mass fraction, which further tightens the lower mass bound to $10^{-16}M_\odot$, yielding the upper bound for the density scale of a positive-valued $q$: 
\begin{equation} \label{qconstraintPBHLim}
   1/|q| \lesssim 10^{33}~\mathrm{GeV}/\mathrm{fm}^3\sim10^{68}~\mathrm{J}/\mathrm{m}^3 \quad \text{for}~ q>0 .
\end{equation}

We reiterate that the lower bounds on black hole mass that we have considered here apply only to the initial mass of black holes formed from the gravitational collapse of matter (described as a perfect fluid subject to monotonicity constraints on the EoS). PBHs formed from other mechanisms, for instance from the collapse of topological defects \cite{Hawking:1987bn} or from gravitational waves \cite{Abrahams:1992ib,Alcubierre:1999ex}, are not necessarily subject to these bounds. It should be emphasized that Fig. \ref{fig:PBHConstraints}, which only considers mass scales above $10^{-18}~M_\odot$, does not constrain the dark matter mass fraction for Planck scale remnants \cite{adler2001generalized,chen2003black} resulting from the evaporation of low-mass PBHs. Moreover, the mechanism proposed here does not prevent a black hole evaporating via Hawking radiation from crossing the threshold in Eq. \eqref{minmass_in_q}.

%
%

%
%
\section{Numerical solutions}\label{sec:Numerical}

The TOV equations for the MEMe model are governed by,

\begin{equation} \label{TOV_MEMe}
   \begin{aligned}
      m'(r) &= 4\pi \rho(r) r^2, \\
      p'(r) &= -[p(r)+\rho(r)]\frac{m(r)+4\pi r^3 p(r)}{[r-2 m(r)]r}, \\
      \rho'(r) &= - \left(\frac{\partial F}{\partial \rho}\right)^{-1} \left(\frac{\partial F}{\partial p}\right)p'(r).
   \end{aligned}
\end{equation}

For a simple EoS~\eqref{linearEOS}, $F(\rho,p)$ is given by~\eqref{effEOS} which is also used to calculate the corresponding value of $\partial F / \partial \rho$ and $\partial F / \partial p$. For the rest of this paper, we will only consider the special case of dust by setting $w$ in~\eqref{linearEOS} to zero.

Given a starting radius of $r_0 \rightarrow 0$, the initial conditions are,
\begin{equation} \label{MEMe_initial}
   \begin{aligned}
      m(r_0) &= \frac{4}{3}\pi \rho_0 r_0 ^3, \\
      \rho(r_0) &= \rho_0, \\
      p(r_0) &= p_0,
   \end{aligned}
\end{equation}
where $p_0$ is obtained by solving \eqref{effEOS} for $\rho_0$ and choosing a solution that crosses $p = \rho = 0$. The chosen initial density $\rho_0$ needs to lie on the implicit EoS, $F(\rho,p)$~\eqref{effEOS}. As shown in the plot of the implicit EoS (Figure \ref{fig:EoS}), there are three regions of interest. The first region is between the origin and point $\textbf{A}$, where $\partial p / \partial \rho >0$ and $\rho >0$. Next, we have the second region between points $\textbf{A}$ and $\textbf{B}$, where $\partial p / \partial \rho <0$ and $\rho >0$. Lastly, the third region is between points $\textbf{B}$ and $\textbf{C}$,  where $\partial p / \partial \rho >0$ and $\rho <0$. We will consider three different scenarios in terms of initial conditions, based on these three regions.

Although we must choose a value for the parameter $q$ to proceed with numerical solutions, we found that the qualitative properties of the solutions are insensitive to variations in $q$. Currently we have set $q \sim 5 \times 10^{-50} ~\mathrm{m^3/J}$; we will further explore the implications of its scale later on.

\begin{figure}
   \includegraphics[width=0.48\textwidth]{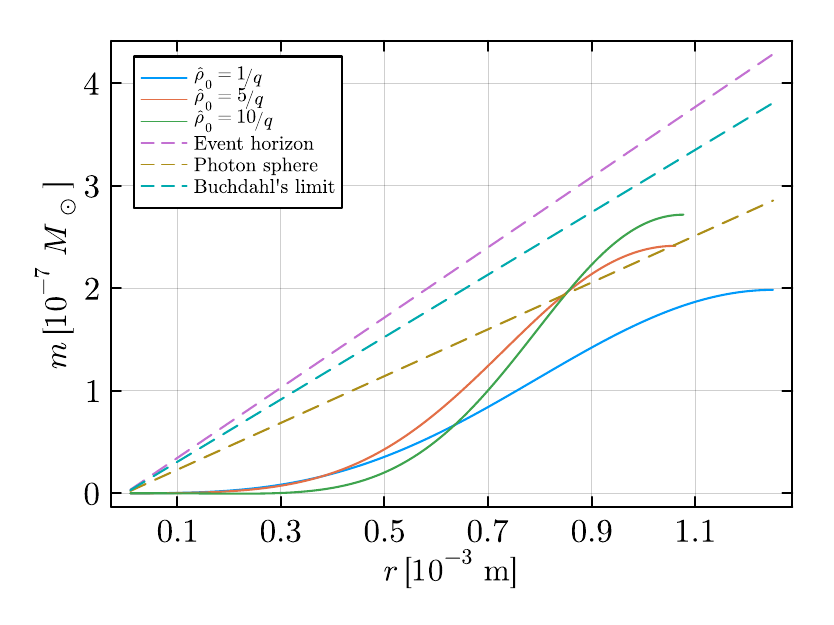}
   \caption{The resulting mass function $m(r)$ for each different scenario, $\hat{\rho}_0 = 1/q,~5/q,~$and $10/q$. $m(r)$ is terminated when $P(r) = 0$. Event horizon~$(r = 2m)$, photon sphere ~$(r = 3m)$, and Buchdahl's limit~$(r = 9m/4)$ are indicated by dashed lines. }
   \label{fig:M}
\end{figure}

We calculate the solutions\footnote{For the numerical integration of the TOV equations, we employ the solvers in the Julia language library OrdinaryDiffEq.jl \cite{SciMLDiffEq:2017}.} based on the three regions in the EoS, each of which has the initial physical density $\hat{\rho}_0 = 1/q,~5/q,~10/q$, respectively. The resulting mass function is given in figure~\ref{fig:M}. The plot is terminated when $p(r) = 0$, signaling the boundary of the matter distribution. The resulting object mass $M$ and radius $R$ are indicated by this end point. We find that a compact object solution exists in all regions. Particularly for the second and third regions, we find the resulting object's radius lies near and even smaller than their corresponding photon sphere  ($r = 3 m$). 

\begin{figure}
   \includegraphics[width=0.48\textwidth]{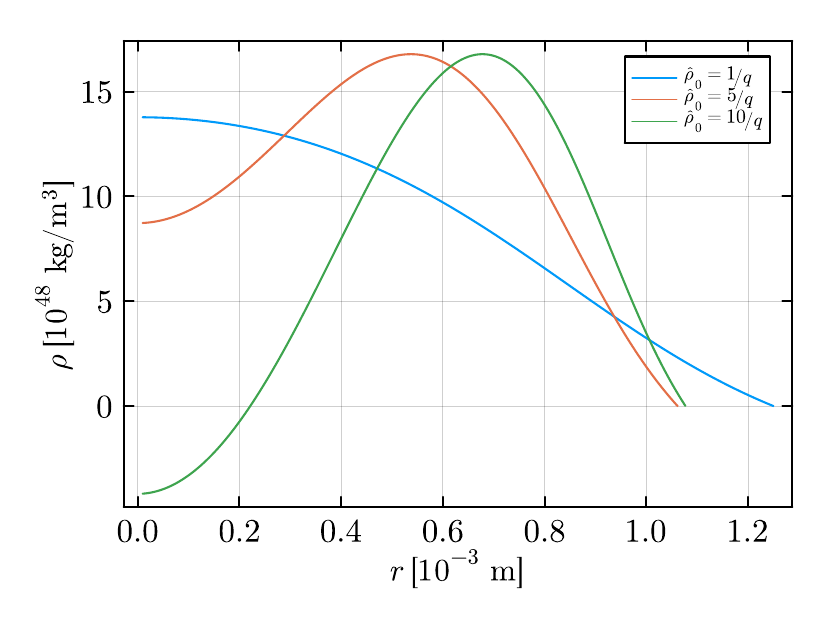}
   \caption{The density function $\rho(r)$ for each different scenario, $\hat{\rho}_0 = 1/q,~5/q,~$and $10/q$. Only the first scenario has $\rho(r)$ that is monotonically decreasing. The other scenarios started with an increasing~$\rho(r)$ before they peaked at $\rho_{\rm max}$ and flipped to a decreasing function. Additionally the third scenario started from a negative density in the Einstein frame.}
   \label{fig:rho}
\end{figure}

One can find an explanation for this behavior by looking into the Einstein frame density profiles of the solutions, presented in the figure~\ref{fig:rho}. For the solution from the first region, the density profile is monotonically decreasing, as what we normally expect. Meanwhile for the second and third regions, $\rho(r)$ will initially increase from the center $r = 0$ to a point where the maximum Einstein frame density $\rho_{\rm max}$ is reached, then it will proceed to decrease to zero at the end. In the third region, this behavior is more pronounced, as the density now starts from a negative Einstein frame value at the center. This leads to the existence of a negative mass distribution at the object’s center, surrounded by a shell of positive mass distribution. While the existence of negative Einstein frame density $\rho$ and negative mass distribution may seem worrisome, it should be stressed that this is due to our transformation to the Einstein frame. In the physical Jordan frame, the physical density $\hat{\rho}$ is always positive and monotonically decreasing for all solutions.

\begin{figure}
   \includegraphics[width=0.48\textwidth]{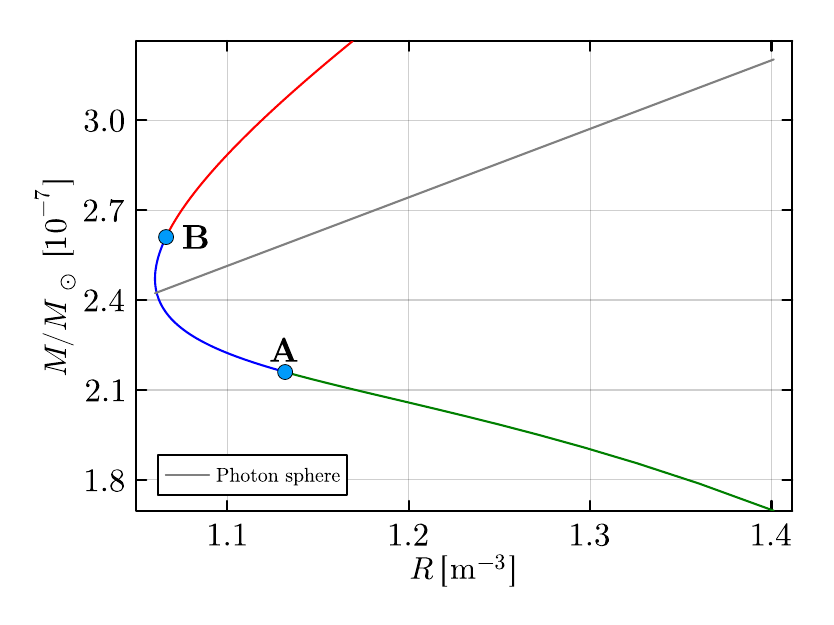}
   \caption{The resulting object mass $M$ and radius $R$ curve, with various value of initial density. The first, second, and third regions of figure~\ref{fig:EoS} are indicated by green, blue, and red curve, respectively. Points $\textbf{A}$ and $\textbf{B}$ from figure~\ref{fig:EoS} are also shown. Additionally photon sphere  line is shown in a solid gray line.}
   \label{fig:MR}
\end{figure}

Based on the resulting object mass $M$ and radius $R$, we produce the $M-R$ curve for the MEMe model by varying the initial physical density $\hat{\rho}_0$, ranging from $1/2q <\hat{\rho}_0 < 30/q$. In figure \ref{fig:MR}, the first, second, and third regions are indicated by the green, blue, and red curves, respectively. The lowest $\hat{\rho}_0$ is located at the bottom right of the figure. As we increase $\hat{\rho}_0$, the resulting object mass $M$ increases while its radius $R$ decreases. This behavior persists as we cross point $\textbf{A}$, moving into the second region. Here $R$ decreases slower as we increase $\hat{\rho}_0$. At some point, this behavior is flipped, and $R$ increases as we increase $\hat{\rho}_0$. Then in the third region, this increase of $R$ continues. For all regions, $M$ will always increase as we increase $\hat{\rho}_0$.

\begin{figure}
   \includegraphics[width=0.48\textwidth]{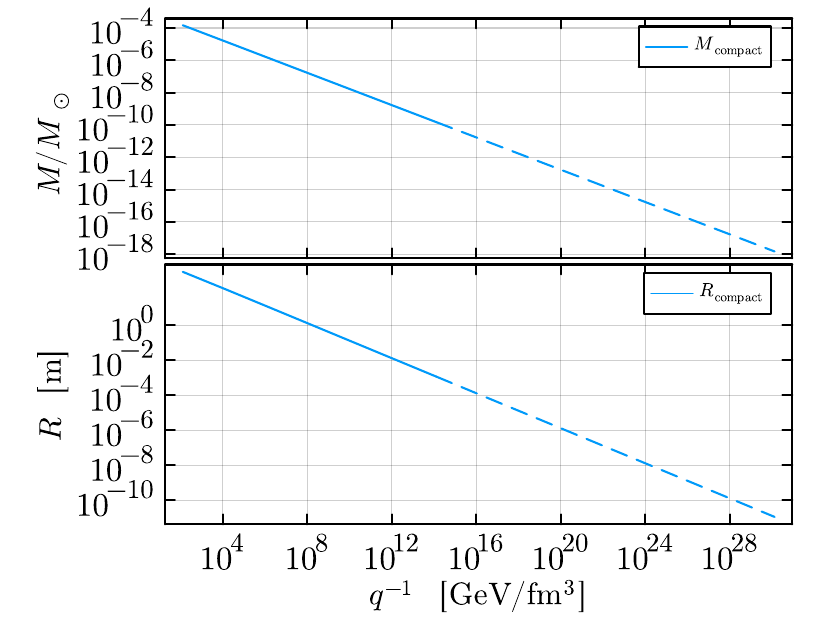}
   \caption{The mass $M_{\text{compact}}$ as a function of $q^{-1}$ is shown in the upper plot, while the radius $R_{\text{compact}}$ is displayed in the lower plot, both assuming a central density $\hat{\rho}_0 = 1/q$. Solid blue lines represent numerical results, and dashed lines indicate extrapolated values.}
   \label{fig:MRvsq_compact}
\end{figure}

Returning to the discussion about parameter $q$, we illustrate the relation between the resulting object properties, $M$ and $R$, with $\mu$ in figure~\ref{fig:MRvsq_compact}. As discussed in the previous section, one can connect $q$ with $\mu$, the energy scale where the effective geometry breaks down. Here we choose $\hat{\rho}_0 = 1/q$ as the central density.
We numerically calculate the resulting object $M$ and $R$ for various $\mu$ and put it on figure~\ref{fig:MRvsq_compact} with the blue line, while the dashed line indicates a result obtained by using an extrapolation. 
Here $M$ and $R$ scale with $\mu$ in the same way.
We find that the object mass $M_{\rm compact} (\mu)$ is smaller from the minimum black hole mass $M_{\rm min} (\mu)$ for the corresponding energy scale $\mu$ by three orders of magnitude. Meanwhile for the object radius $R_{\rm compact} (\mu)$, it is on the order of, though slightly larger than, the minimum black hole radius for the same energy scale $\mu$.

\section{Stability Conditions}\label{sec:stabcond}

When considering the stability of compact objects, the typical condition for the one-phase compact object is $\partial M / \partial \rho_0 > 0$. We can see from figure \ref{fig:MR} that this condition is satisfied for all ranges of $\rho_0$. Considering the nature of our EoS, it is difficult to accept this condition as it is; in particular, the Einstein frame EoS permits negative energy densities that are bounded only by the critical density, which may signal the presence of an instability. As such it is necessary to do a more thoughtful examination by the mean of radial perturbations~\cite{Pretel:2020xuo}.

Given an equilibrium configuration from the TOV equations~\eqref{TOV0}, we consider a small deviation of this equilibrium configuration while still keeping the spherical symmetry. This is achieved by solving the perturbed Einstein equations $\delta G_{\mu\nu} = \kappa \delta T_{\mu\nu}$ for small radial oscillations. The radial perturbation will displace a fluid element at coordinate $r$ in the unperturbed configuration to radial coordinate $r + \xi(t,r)$ in the perturbed configuration, where $\xi(t,r)$ is the Lagrangian displacement. To solve the perturbed Einstein equations, it is important to make a clear distinction between two different perspectives of a fluid perturbation~\cite{Shapiro:1983du}. The first is the Eulerian perturbation, which for some fluid function $Q$ is defined as $\delta Q (t,r) = Q(t,r) - Q_{\mathrm{i}}(r)$. Here the subscript ${\mathrm{i}}$ denotes the unperturbed configuration. One can think of the Eulerian perturbation as measuring the change for a fixed point in the coordinate system. On the other hand, the Lagrangian perturbation is defined as $\Delta Q(t,r) = Q(t, r + \xi(t,r)) - Q_{\mathrm{i}}(r)$. In this perspective, one measures the change of fluid function from an observer that moves alongside the fluid.

The task of reducing the perturbed Einstein equations has been done extensively by S. Chandrasekhar~\cite{Chandrasekhar:1964zz,Chandrasekhar:1964zza}, and G. Chanmugam~\cite{chanmugam1977radial}. One important assumption is that all perturbations are assumed to have a harmonic time dependence, i.e. $\Delta Q(t,r) = \Delta Q(r)e^{i\omega t}$, where $\omega$ is the characteristic oscillation frequency. The oscillation equations are now reduced to two differential equations in term of the perturbation amplitudes and functions of the equilibrium configuration. Defining a new function $\zeta \equiv \xi/r$, the radial oscillation equations are given by,

\begin{align}
    \frac{d\zeta}{dr}  &= - \left(\frac{3}{r} + \frac{dp}{dr} \frac{1}{p + \rho} \right) \zeta -\frac{1}{r} \frac{\Delta p}{\Gamma p}, \label{eq:SL1}\\
    \frac{d(\Delta p)}{dr}  &= \zeta \left\{ \frac{(p + \rho)r \omega^2}{e^{2\phi} (1-2m/r)} -4 \frac{dp}{dr}\right\} \nonumber\\
    & + \zeta \left\{ \left( \frac{dp}{dr}\right)^2 \frac{r}{(p+\rho)} -\frac{8\pi(p+\rho) p r}{(1-2m/r)} \right\}\nonumber\\
    &+ \Delta p \left\{ \frac{dp}{dr} \frac{1}{p+\rho} - \frac{4\pi (p+\rho)r}{(1-2m/r)} \right\} \label{eq:SL2}.
\end{align}

Here besides the perturbation amplitudes $\zeta$ and $\Delta p$, all functions are now considered in the unperturbed configuration. The function $\Gamma$ is the adiabatic index governing the perturbations at a constant entropy. Given an EoS in the form of $P = P(\rho) $, one can obtain the adiabatic index by considering the variation of $p$. Hence, the adiabatic index is given by, 

\begin{equation}
    \Gamma(r) = \left(1 + \frac{\rho}{p}\right)\frac{d p}{d\rho}.
\end{equation}

The same result will be achieved if we start with an implicit form of EoS, $F(\rho, p) = 0$.

When combined, Eqs. \eqref{eq:SL1} and \eqref{eq:SL2} have the form of a Sturm-Liouville type problem for the radial oscillation modes, and one might expect the modes to have a discrete spectrum. However, we caution that Eq. \eqref{eq:SL1} becomes singular at $r=0$, and one must assume that the singular point does not spoil the discreteness of the spectrum---we shall proceed under this assumption and follow the standard analysis, but it is important to keep in mind that this analysis is heuristic and that a more rigorous approach is needed to conclusively establish the radial stability of solutions to the TOV equation \cite{Luz:2024lgi,Luz:2024xnd,Luz:2024yjm}.

We assume the radial oscillations are characterized by discrete eigenvalues 
$\omega_n^2$ (which can have negative values), where $n\in\mathbb{Z}$. 
A given equilibrium configuration is unstable if for any $n$, $\omega_n^2 <0$, as this will lead to a growing mode. To solve equations~\eqref{eq:SL1} and~\eqref{eq:SL2} numerically, it is required to provide the appropriate boundary conditions. At the center, $r \rightarrow 0$, the Eq.~\eqref{eq:SL1} contains a singularity. To ensure the solutions are regular, the coefficient $1/r$ must vanish as $r\rightarrow 0$, hence
\begin{equation}
    \Delta p = -3\Gamma \zeta p \quad \mathrm{as} \quad r\rightarrow 0. \label{eq:BC1}
\end{equation}

At the surface of the object, $p(R) = 0$. As such it is appropriate for the Lagrangian perturbation of pressure to also vanish, 
\begin{equation}
    \Delta p = 0 \quad r\rightarrow R. \label{eq:BC2}
\end{equation}

Due to the spherical symmetry, the necessary boundary condition for Lagrangian displacement is $\xi(0) = 0$. But as our equations are given in terms of $\zeta$, it is appropriate to normalize this function so $\zeta(0) = 1$ at the center.

Using an equilibrium solution of the MEMe model, which is discussed in section~\ref{sec:Numerical}, one can solve the radial oscillation equations, Eq. ~\eqref{eq:SL1} and~\eqref{eq:SL2}, to check the stability of such solution. Using the Einstein frame, the modification of gravity is now encoded in the form of the Einstein frame pressure $p$ and density $\rho$, making the radial oscillation equations can be used without any modification. One further requirement is to provide the metric function $\phi(r)$, which can be directly calculated from available functions by using Eq.~\eqref{EinsteinSph} and matching the boundary condition to match a Schwarzschild solution at the surface $r = R$.

We numerically solve the Eq.~\eqref{eq:SL1} and~\eqref{eq:SL2} by using the shooting method. First, we need to specify the TOV solution that we want to examine. Afterward, we integrate for various trial values of $\omega^2$ with the initial conditions set by the boundary condition~\eqref{eq:BC1} and the normalized $\zeta(0) = 1$ condition. After integrating it to the surface $R$, the values of $\omega^2$ that satisfy the second boundary condition~\eqref{eq:BC2} are the correct solutions. In principle, we will have several eigenvalues $\omega_n^2$, which will have their own corresponding $\zeta_n(r)$ and $\Delta p_n(r)$, where $n$ indicates the number of nodes inside the object. The fundamental oscillation mode is identified by looking at the solution that has no node. 

\begin{figure}
   \includegraphics[width=0.48\textwidth]{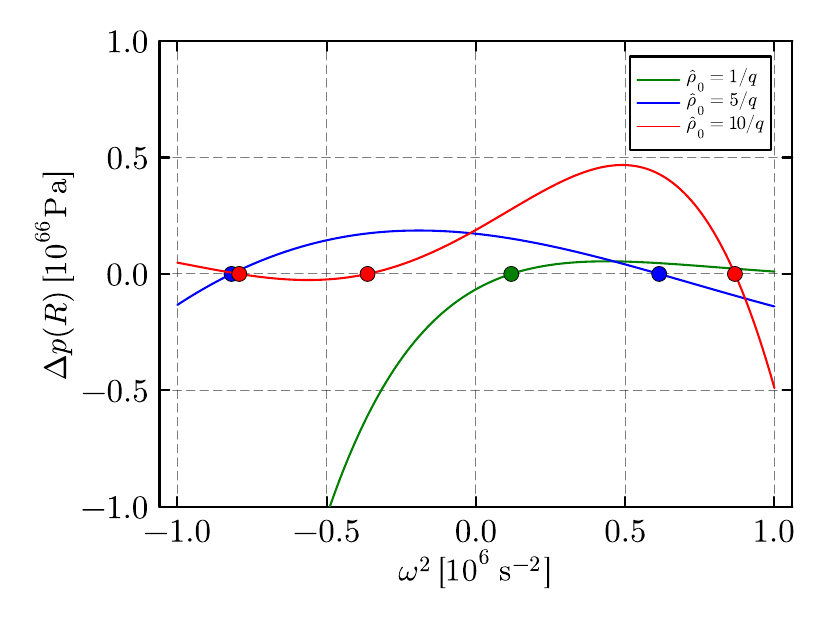}
   \caption{The Lagrangian perturbation of pressure at the surface, $\Delta p(R)$, as a function of the trial eigenvalues $\omega^2$ for three TOV solutions with initial conditions $\hat{\rho}_0 = 1/q$, $5/q$, and $10/q$, represented by the green, blue, and red curves, respectively. The eigenvalues, corresponding to the points where $\Delta p(R) = 0$, are indicated with markers.}
   \label{fig:delp_omega}
\end{figure}

We start our calculation from $\omega^2 = -1 \times10^6 ~\rm s^{-2}$, then we continue the calculation until we have three solutions. We do a separate calculation for different TOV solutions, specified by their initial condition $\hat{\rho}_0 = 1/q,~5/q,$ and $~10/q$. Their subset of eigenvalues can be found in the table~\ref{tab:omega2}. Alternatively, we plot the Lagrangian perturbation of pressure at the surface, $\Delta p (R)$, for these TOV solutions as a function of trial values $\omega^2$ in figure~\ref{fig:delp_omega}. Here the eigenvalues are indicated at the points where $\Delta p (R) = 0$. We find that the only TOV solution that does not have the negative $\omega^2_n$ is the one with the initial condition in the first region of EoS, $\hat{\rho}_0 = 1/q$, while others have at least one negative $\omega^2_n$. We can conclude that these other TOV solutions, $\hat{\rho}_0 = 5/q$ and $\hat{\rho}_0 = 10/q$, are unstable due to the existence of the negative $\omega^2_n$.

\begin{figure}
   \includegraphics[width=0.48\textwidth]{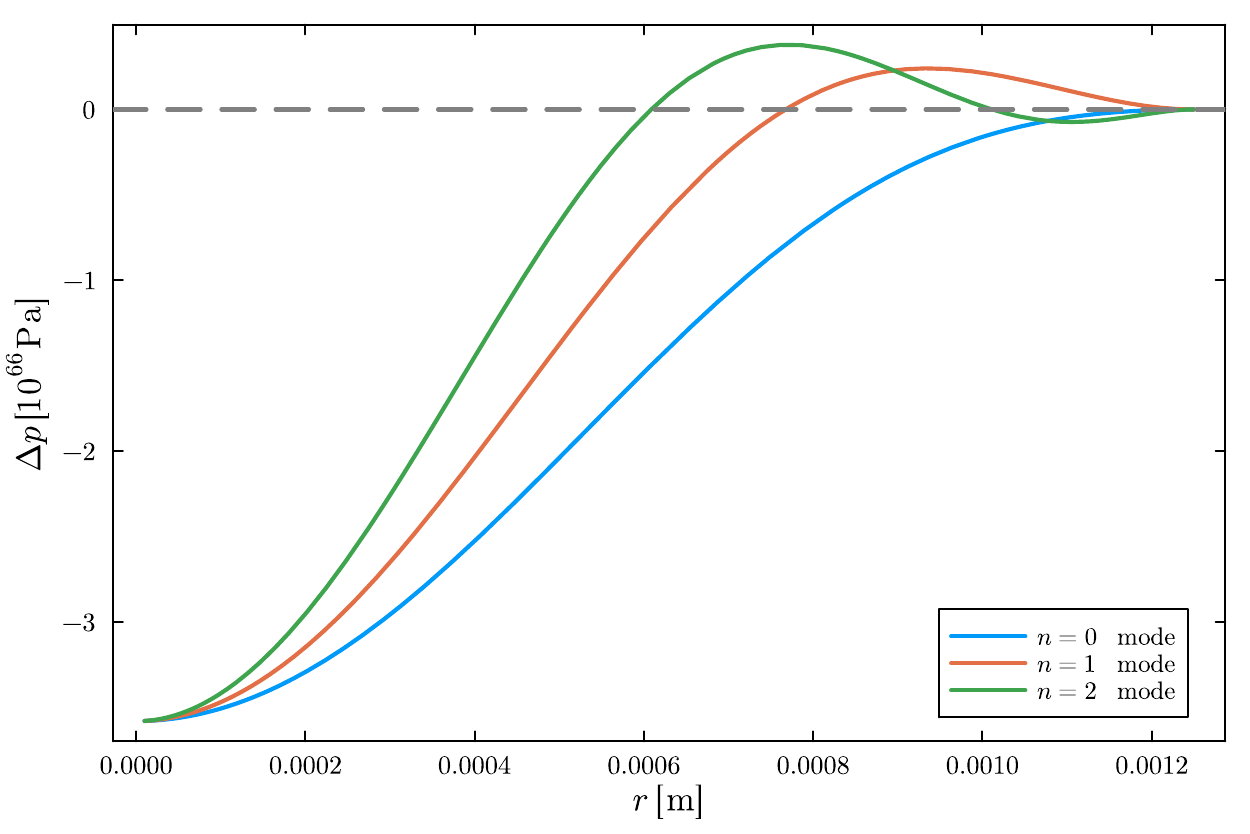}
   \caption{The Lagrangian perturbation of pressure, $\Delta p(r)$, as a function of radius $r$ for the fundamental mode ($n=0$, blue curve) and the first two excited modes ($n=1$, orange curve; $n=2$, green curve). The fundamental mode is identified by $\Delta p(r)$ remaining non-negative and crossing zero only once at the surface, while the higher modes cross zero multiple times within the star's radius.}
   \label{fig:delp}
\end{figure}

For the first region TOV solution, we need to check whether any eigenvalue that we found is indeed the fundamental mode. We plot the Lagrangian perturbation of pressure $\Delta p (r)$ for these three eigenvalues in the figure~\ref{fig:delp}. We find that $\Delta p(r)$ corresponding to the lowest $\omega^2$ never crossed zero between the boundaries, hence it can identified as the fundamental mode. The $\Delta p (r)$ for the other eigenvalues are also shown, and they can be identified as $n = 1$ and $n = 2$ modes. As the fundamental eigenvalue $\omega^2_n$ is positive, we can safely conclude that this TOV solution is stable. Furthermore, we have checked that this condition is satisfied not only for the TOV solution with initial condition $\hat{\rho}_0 = 1/q$, but also all TOV solutions in the first region of EoS, $\hat{\rho}_0 < 2/q$.

\begin{table}[]
\begin{tabular}{|l|l|l|l|}
\hline
$\hat{\rho}_0$ & $\omega_n^2~[10^6~ \rm s^{-2}]$ \\ \hline
$1/q$   &  $[0.118502,~1.12738,~2.52638]$ \\ \hline
$5/q$   & $[-0.818704,~0.614103,~2.81277]$  \\ \hline
$10/q$   & $[-0.792770,~-0.362850,~0.867784]$   \\ \hline
\end{tabular}
\caption{A subset of eigenvalues $\omega_n^2$ for different TOV solutions, characterized by their initial condition $\hat{\rho}_0 = 1/q$, $5/q$, and $10/q$. The eigenvalues are obtained from the stability analysis, where negative values of $\omega_n^2$ indicate instability in the corresponding TOV solution.}
\label{tab:omega2}
\end{table}

%
%
%
%

\section{Summary and further discussion}\label{sec:disc}
In this work, we investigated the phenomenology of compact object solutions arising from an Equation of State (EoS) for dust, within the MEMe model of gravity. Such objects are permitted when the single free parameter $q$ of the MEMe model is positive.
We obtained the compact objects by numerically solving the Tolman-Oppenheimer-Volkov (TOV) equations~\eqref{TOV_MEMe} for the Einstein-frame pressure and density. 
Our analysis reveals that the compact object's mass $M$ and radius $R$ both scale as $1/q$, $1/q$ representing the energy density scale at which the effective geometry description ceases to be valid. The resulting solutions can be classified into three distinct classes according to their initial central density values, $\hat{\rho}_0$. Furthermore, we conducted a stability analysis through radial perturbations of each solution class. We find that only the first class of solutions, characterized by an initial central pressure (in the Einstein frame) that is less than the pressure corresponding to a turning point density $\rho_{\rm max}$, are stable to radial perturbations. The other classes of solutions were found to be unstable, despite potentially describing ultra-compact objects with radii smaller than their respective photon spheres.

We considered the general behavior of the Einstein frame density for matter with a monotonically increasing physical EoS, and found that for positive $q$, the Einstein frame density has an upper bound uniquely saturated by dust, coinciding with the turning point density $\rho_{\rm max}$. This yields an expression for the corresponding lower bound for the mass of black holes formed from the gravitational collapse of matter, which we have derived in Eq. \eqref{minmass_in_q}. If we postulate that the lack of observed signals from the late stages of PBH evaporation is a consequence of such a lower bound, then we can use observational constraints on Primordial Black Hole (PBH) mass (as indicated in Fig. \ref{fig:PBHConstraints} and \cite{Carr:2020gox}) to postulate a lower bound on positive values for the parameter $q$ in Eq. \eqref{qconstraintPBHLim}. This scenario provides an relatively simple alternative to other proposed explanations \cite{Santiago:2025rzb,*Baker:2025zxm,Davies:2024ysj,Carr:2021bzv,Bai:2019zcd,Chen:2002tu,Adler:2001vs,Barrow:1992hq,Barrow:1992ay,Aharonov:1987tp,Dvali:2025ktz,*Thoss:2024hsr,*Dvali:2024hsb,*Alexandre:2024nuo,*Dvali:2020wft,*Dvali:2018xpy} for the lack of observed Hawking radiation from low-mass PBHs, provided that $q>0$; the maximum Einstein frame density in the MEMe model prevents the formation of black holes below a certain mass threshold in the first place.

In addition to constraining the masses of PBHs, microlensing studies \cite{MACHO:2000qbb,Basak:2021ten,Sammons:2020kyk} indicated in Fig. \ref{fig:PBHConstraints} (combined with supernova samples \cite{DES:2024ffp} and dynamical heating considerations \cite{Brandt:2016aco}) constrain Massive Compact Halo Objects (MACHOs) for a wide range of mass scales. The dust star solutions we have found are a class of MACHOs, and are horizonless compact objects that can in some respects mimic the behavior of PBHs. Our dust star solutions can for certain values of $q$ comprise a significant fraction of dark matter; a comparison of the mass ranges identified in Fig. \ref{fig:MRvsq_compact} with the microlensing constraints in Fig. \ref{fig:PBHConstraints} indicate that if $1/|q|\gtrsim 10^{18}~\mathrm{GeV}/\mathrm{fm}^3$, corresponding to a mass $10^{-11}M_\odot$, then the dust stars can evade existing mass constraints on MACHOs. Since this is less than the upper limit $1/|q| \lesssim 10^{33}~\mathrm{GeV}/\mathrm{fm}^3$ given in Eq. \eqref{qconstraintPBHLim}, one may conclude that there is a range of scales for $q>0$ such that PBH formation in this sector is suppressed below $10^{-16}~M_\odot$ and in which the dust star solutions we considered in this article can (in addition to PBHs) form a significant fraction of dark matter.

A natural next step will be to extend the present analysis to matter comprised of scalar and vector fields. So far, a detailed investigation of the MEMe model for matter sources beyond that of a perfect fluid has yet to be performed. Such an analysis is needed to establish the effects of the modified gravitational coupling on the properties of boson stars \cite{Liebling:2012fv,Visinelli:2021uve,Kaup:1968}.

Another important question for future consideration concerns the details of PBH formation (see \cite{LISACosmologyWorkingGroup:2023njw} for a recent review) as well as the compatibility of the scenario considered here with other models for the early universe. In particular, a key question of interest is whether our considerations, which require $q>0$, can be made compatible with the bounce solutions found in \cite{Troisi:2025ohh}, which require $q<0$. In the appendix, we constructed a multi-MEMe model, which includes sectors in which $q<0$ and in which $q>0$, though a more detailed study will be required to determine whether such an extended MEMe model can accommodate both the PBH suppression mechanism and a bounce within a single framework. Even in the absence of such a framework, the suppression mechanism presented here may inspire the formulation of alternative mechanisms for suppressing the formation of PBHs below a certain mass scale.


%
%


\begin{acknowledgments}
\noindent
We thank Sante Carloni for informative discussions, and for making us aware of recent work with Paulo Luz on the radial stability of compact objects. We also thank Sebastian Schuster for making us aware of the PlotPBHbounds package. RDL and PC was supported by the Leung Center for Cosmology and Particle Astrophysics (LeCosPA), National Taiwan University (NTU). JCF acknowledges support from the Leung Center for Cosmology and Particle Astrophysics (LeCosPA), National Taiwan University (NTU), the R.O.C. (Taiwan) National Science and Technology Council (NSTC) Grant No. 112-2811-M-002-132, and the European Union and Czech Ministry of Education, Youth and Sports through the FORTE project No. CZ.02.01.01/00/22\_008/0004632.
\end{acknowledgments}

%
%

\appendix*
\section{Multi-MEMe model}
In this appendix, we consider an extension of the MEMe model to
accommodate both positive and negative values for $q$. We introduce the multi-MEMe model, which has a Lagrangian of the form:
\begin{equation}\label{GCA-MultiMEMeAction}
   \begin{aligned}
   S[\Phi,g^{\cdot\cdot},A{_{\cdot}}{^{\cdot}}]= \int & d^4x \biggl\{ \frac{1}{2 \kappa}\left[R - 2 \, \tilde{\Lambda} \right]\sqrt{-{g}} \\
   &+ \sum_i \left[L_{(\mathrm{i})}[\Phi_{(\mathrm{i})},\bar{\mathfrak{g}}_{(\mathrm{i})}^{\cdot\cdot}] - \frac{\lambda_{(\mathrm{i})}}{\kappa} \right] \sqrt{-\mathfrak{g}_{(\mathrm{i})}} \biggr\} ,
   \end{aligned}
\end{equation}
where $\bar{\mathfrak{g}}_{(0)}^{\mu\nu}:=g^{\mu\nu}$, and $\bar{\mathfrak{g}}_{(\mathrm{i})}^{\mu\nu}$ and ${\mathfrak{g}}^{(\mathrm{i})}_{\mu\nu}$ are constructed from the tensors ${A}{^{(\mathrm{i})}}{^{\mu}}_{\alpha}$ and $\bar{A}{_{(\mathrm{i})}}^{\alpha}{_{\mu}}$. The index $i\in \mathbb{N}$ distinguishes sectors with a different value of $\lambda=\lambda_{(\mathrm{i})}$. The variation of the action \eqref{GCA-MultiMEMeAction} with respect to ${A}{^{(\mathrm{i})}}{^{\mu}}_{\alpha}$ yields multiple equations of the form in Eq. \eqref{GCA-ExpFEs} and a single equation generalizing Eq. \eqref{GCA-GEN-GFE-MEMe}. If each sector can be modeled independently as a perfect fluid, then one can solve for the tensors ${A}{^{(\mathrm{i})}}{^{\mu}}_{\alpha}$ independently, then rewrite the gravitational equation as a multifluid Einstein equation, with the effects from the multi-MEMe model encapsulated in modifications to the EoS. One may observe that a consequence of the multi-MEMe model is that the equivalence principle is necessarily relaxed. One might, for instance, require that standard model degrees of freedom are coupled to the same effective metric, and postulate that the dark sector degrees of freedom are coupled to a different effective metric. The multi-MEMe model also offers a pathway for avoiding the fine-tuning problem for the cosmological constant; one can postulate a symmetry in the matter sector (a supersymmetry, for instance) that renders to zero the sum of all $\lambda_{(\mathrm{i})}$.


\bibliography{refs}

\begin{thebibliography}{67}%
\makeatletter
\providecommand \@ifxundefined [1]{%
 \@ifx{#1\undefined}
}%
\providecommand \@ifnum [1]{%
 \ifnum #1\expandafter \@firstoftwo
 \else \expandafter \@secondoftwo
 \fi
}%
\providecommand \@ifx [1]{%
 \ifx #1\expandafter \@firstoftwo
 \else \expandafter \@secondoftwo
 \fi
}%
\providecommand \natexlab [1]{#1}%
\providecommand \enquote  [1]{``#1''}%
\providecommand \bibnamefont  [1]{#1}%
\providecommand \bibfnamefont [1]{#1}%
\providecommand \citenamefont [1]{#1}%
\providecommand \href@noop [0]{\@secondoftwo}%
\providecommand \href [0]{\begingroup \@sanitize@url \@href}%
\providecommand \@href[1]{\@@startlink{#1}\@@href}%
\providecommand \@@href[1]{\endgroup#1\@@endlink}%
\providecommand \@sanitize@url [0]{\catcode `\\12\catcode `\$12\catcode `\&12\catcode `\#12\catcode `\^12\catcode `\_12\catcode `\%12\relax}%
\providecommand \@@startlink[1]{}%
\providecommand \@@endlink[0]{}%
\providecommand \url  [0]{\begingroup\@sanitize@url \@url }%
\providecommand \@url [1]{\endgroup\@href {#1}{\urlprefix }}%
\providecommand \urlprefix  [0]{URL }%
\providecommand \Eprint [0]{\href }%
\providecommand \doibase [0]{http://dx.doi.org/}%
\providecommand \selectlanguage [0]{\@gobble}%
\providecommand \bibinfo  [0]{\@secondoftwo}%
\providecommand \bibfield  [0]{\@secondoftwo}%
\providecommand \translation [1]{[#1]}%
\providecommand \BibitemOpen [0]{}%
\providecommand \bibitemStop [0]{}%
\providecommand \bibitemNoStop [0]{.\EOS\space}%
\providecommand \EOS [0]{\spacefactor3000\relax}%
\providecommand \BibitemShut  [1]{\csname bibitem#1\endcsname}%
\let\auto@bib@innerbib\@empty
\bibitem [{\citenamefont {Cardoso}\ and\ \citenamefont {Pani}(2017)}]{Cardoso:2017njb}%
  \BibitemOpen
  \bibfield  {author} {\bibinfo {author} {\bibfnamefont {V.}~\bibnamefont {Cardoso}}\ and\ \bibinfo {author} {\bibfnamefont {P.}~\bibnamefont {Pani}},\ }\href@noop {} {\  (\bibinfo {year} {2017})},\ \Eprint {http://arxiv.org/abs/1707.03021} {arXiv:1707.03021 [gr-qc]} \BibitemShut {NoStop}%
\bibitem [{\citenamefont {{Goebel}}(1972)}]{Goebel:1972}%
  \BibitemOpen
  \bibfield  {author} {\bibinfo {author} {\bibfnamefont {C.~J.}\ \bibnamefont {{Goebel}}},\ }\href {\doibase 10.1086/180898} {\bibfield  {journal} {\bibinfo  {journal} {ApJL}\ }\textbf {\bibinfo {volume} {172}},\ \bibinfo {pages} {L95} (\bibinfo {year} {1972})}\BibitemShut {NoStop}%
\bibitem [{\citenamefont {Cardoso}\ \emph {et~al.}(2009)\citenamefont {Cardoso}, \citenamefont {Miranda}, \citenamefont {Berti}, \citenamefont {Witek},\ and\ \citenamefont {Zanchin}}]{Cardoso:2008bp}%
  \BibitemOpen
  \bibfield  {author} {\bibinfo {author} {\bibfnamefont {V.}~\bibnamefont {Cardoso}}, \bibinfo {author} {\bibfnamefont {A.~S.}\ \bibnamefont {Miranda}}, \bibinfo {author} {\bibfnamefont {E.}~\bibnamefont {Berti}}, \bibinfo {author} {\bibfnamefont {H.}~\bibnamefont {Witek}}, \ and\ \bibinfo {author} {\bibfnamefont {V.~T.}\ \bibnamefont {Zanchin}},\ }\href {\doibase 10.1103/PhysRevD.79.064016} {\bibfield  {journal} {\bibinfo  {journal} {Phys. Rev. D}\ }\textbf {\bibinfo {volume} {79}},\ \bibinfo {pages} {064016} (\bibinfo {year} {2009})},\ \Eprint {http://arxiv.org/abs/0812.1806} {arXiv:0812.1806 [hep-th]} \BibitemShut {NoStop}%
\bibitem [{\citenamefont {Berti}(2014)}]{Berti:2014bla}%
  \BibitemOpen
  \bibfield  {author} {\bibinfo {author} {\bibfnamefont {E.}~\bibnamefont {Berti}}\ }(\bibinfo {year} {2014})\ \Eprint {http://arxiv.org/abs/1410.4481} {arXiv:1410.4481 [gr-qc]} \BibitemShut {NoStop}%
\bibitem [{\citenamefont {Glampedakis}\ \emph {et~al.}(2017)\citenamefont {Glampedakis}, \citenamefont {Pappas}, \citenamefont {Silva},\ and\ \citenamefont {Berti}}]{Glampedakis:2017dvb}%
  \BibitemOpen
  \bibfield  {author} {\bibinfo {author} {\bibfnamefont {K.}~\bibnamefont {Glampedakis}}, \bibinfo {author} {\bibfnamefont {G.}~\bibnamefont {Pappas}}, \bibinfo {author} {\bibfnamefont {H.~O.}\ \bibnamefont {Silva}}, \ and\ \bibinfo {author} {\bibfnamefont {E.}~\bibnamefont {Berti}},\ }\href {\doibase 10.1103/PhysRevD.96.064054} {\bibfield  {journal} {\bibinfo  {journal} {Phys. Rev. D}\ }\textbf {\bibinfo {volume} {96}},\ \bibinfo {pages} {064054} (\bibinfo {year} {2017})},\ \Eprint {http://arxiv.org/abs/1706.07658} {arXiv:1706.07658 [gr-qc]} \BibitemShut {NoStop}%
\bibitem [{\citenamefont {Shankaranarayanan}\ and\ \citenamefont {Johnson}(2022)}]{Shankaranarayanan:2022wbx}%
  \BibitemOpen
  \bibfield  {author} {\bibinfo {author} {\bibfnamefont {S.}~\bibnamefont {Shankaranarayanan}}\ and\ \bibinfo {author} {\bibfnamefont {J.~P.}\ \bibnamefont {Johnson}},\ }\href {\doibase 10.1007/s10714-022-02927-2} {\bibfield  {journal} {\bibinfo  {journal} {Gen. Rel. Grav.}\ }\textbf {\bibinfo {volume} {54}},\ \bibinfo {pages} {44} (\bibinfo {year} {2022})},\ \Eprint {http://arxiv.org/abs/2204.06533} {arXiv:2204.06533 [gr-qc]} \BibitemShut {NoStop}%
\bibitem [{\citenamefont {Lin}\ and\ \citenamefont {Mukohyama}(2017)}]{Lin:2017oow}%
  \BibitemOpen
  \bibfield  {author} {\bibinfo {author} {\bibfnamefont {C.}~\bibnamefont {Lin}}\ and\ \bibinfo {author} {\bibfnamefont {S.}~\bibnamefont {Mukohyama}},\ }\href {\doibase 10.1088/1475-7516/2017/10/033} {\bibfield  {journal} {\bibinfo  {journal} {JCAP}\ }\textbf {\bibinfo {volume} {10}},\ \bibinfo {pages} {033} (\bibinfo {year} {2017})},\ \Eprint {http://arxiv.org/abs/1708.03757} {arXiv:1708.03757 [gr-qc]} \BibitemShut {NoStop}%
\bibitem [{\citenamefont {De~Felice}\ and\ \citenamefont {Mukohyama}(2021)}]{DeFelice:2020prd}%
  \BibitemOpen
  \bibfield  {author} {\bibinfo {author} {\bibfnamefont {A.}~\bibnamefont {De~Felice}}\ and\ \bibinfo {author} {\bibfnamefont {S.}~\bibnamefont {Mukohyama}},\ }\href {\doibase 10.1088/1475-7516/2021/04/018} {\bibfield  {journal} {\bibinfo  {journal} {JCAP}\ }\textbf {\bibinfo {volume} {04}},\ \bibinfo {pages} {018} (\bibinfo {year} {2021})},\ \Eprint {http://arxiv.org/abs/2011.04188} {arXiv:2011.04188 [astro-ph.CO]} \BibitemShut {NoStop}%
\bibitem [{\citenamefont {De~Felice}\ \emph {et~al.}(2020)\citenamefont {De~Felice}, \citenamefont {Doll},\ and\ \citenamefont {Mukohyama}}]{DeFelice:2020eju}%
  \BibitemOpen
  \bibfield  {author} {\bibinfo {author} {\bibfnamefont {A.}~\bibnamefont {De~Felice}}, \bibinfo {author} {\bibfnamefont {A.}~\bibnamefont {Doll}}, \ and\ \bibinfo {author} {\bibfnamefont {S.}~\bibnamefont {Mukohyama}},\ }\href {\doibase 10.1088/1475-7516/2020/09/034} {\bibfield  {journal} {\bibinfo  {journal} {JCAP}\ }\textbf {\bibinfo {volume} {09}},\ \bibinfo {pages} {034} (\bibinfo {year} {2020})},\ \Eprint {http://arxiv.org/abs/2004.12549} {arXiv:2004.12549 [gr-qc]} \BibitemShut {NoStop}%
\bibitem [{\citenamefont {Aoki}\ \emph {et~al.}(2019)\citenamefont {Aoki}, \citenamefont {De~Felice}, \citenamefont {Lin}, \citenamefont {Mukohyama},\ and\ \citenamefont {Oliosi}}]{Aoki:2018brq}%
  \BibitemOpen
  \bibfield  {author} {\bibinfo {author} {\bibfnamefont {K.}~\bibnamefont {Aoki}}, \bibinfo {author} {\bibfnamefont {A.}~\bibnamefont {De~Felice}}, \bibinfo {author} {\bibfnamefont {C.}~\bibnamefont {Lin}}, \bibinfo {author} {\bibfnamefont {S.}~\bibnamefont {Mukohyama}}, \ and\ \bibinfo {author} {\bibfnamefont {M.}~\bibnamefont {Oliosi}},\ }\href {\doibase 10.1088/1475-7516/2019/01/017} {\bibfield  {journal} {\bibinfo  {journal} {JCAP}\ }\textbf {\bibinfo {volume} {01}},\ \bibinfo {pages} {017} (\bibinfo {year} {2019})},\ \Eprint {http://arxiv.org/abs/1810.01047} {arXiv:1810.01047 [gr-qc]} \BibitemShut {NoStop}%
\bibitem [{\citenamefont {Aoki}\ \emph {et~al.}(2018)\citenamefont {Aoki}, \citenamefont {Lin},\ and\ \citenamefont {Mukohyama}}]{Aoki:2018zcv}%
  \BibitemOpen
  \bibfield  {author} {\bibinfo {author} {\bibfnamefont {K.}~\bibnamefont {Aoki}}, \bibinfo {author} {\bibfnamefont {C.}~\bibnamefont {Lin}}, \ and\ \bibinfo {author} {\bibfnamefont {S.}~\bibnamefont {Mukohyama}},\ }\href {\doibase 10.1103/PhysRevD.98.044022} {\bibfield  {journal} {\bibinfo  {journal} {Phys. Rev. D}\ }\textbf {\bibinfo {volume} {98}},\ \bibinfo {pages} {044022} (\bibinfo {year} {2018})},\ \Eprint {http://arxiv.org/abs/1804.03902} {arXiv:1804.03902 [gr-qc]} \BibitemShut {NoStop}%
\bibitem [{\citenamefont {Mukohyama}\ and\ \citenamefont {Noui}(2019)}]{Mukohyama:2019unx}%
  \BibitemOpen
  \bibfield  {author} {\bibinfo {author} {\bibfnamefont {S.}~\bibnamefont {Mukohyama}}\ and\ \bibinfo {author} {\bibfnamefont {K.}~\bibnamefont {Noui}},\ }\href {\doibase 10.1088/1475-7516/2019/07/049} {\bibfield  {journal} {\bibinfo  {journal} {JCAP}\ }\textbf {\bibinfo {volume} {07}},\ \bibinfo {pages} {049} (\bibinfo {year} {2019})},\ \Eprint {http://arxiv.org/abs/1905.02000} {arXiv:1905.02000 [gr-qc]} \BibitemShut {NoStop}%
\bibitem [{\citenamefont {Aoki}\ \emph {et~al.}(2020)\citenamefont {Aoki}, \citenamefont {De~Felice}, \citenamefont {Mukohyama}, \citenamefont {Noui}, \citenamefont {Oliosi},\ and\ \citenamefont {Pookkillath}}]{Aoki:2020oqc}%
  \BibitemOpen
  \bibfield  {author} {\bibinfo {author} {\bibfnamefont {K.}~\bibnamefont {Aoki}}, \bibinfo {author} {\bibfnamefont {A.}~\bibnamefont {De~Felice}}, \bibinfo {author} {\bibfnamefont {S.}~\bibnamefont {Mukohyama}}, \bibinfo {author} {\bibfnamefont {K.}~\bibnamefont {Noui}}, \bibinfo {author} {\bibfnamefont {M.}~\bibnamefont {Oliosi}}, \ and\ \bibinfo {author} {\bibfnamefont {M.~C.}\ \bibnamefont {Pookkillath}},\ }\href {\doibase 10.1140/epjc/s10052-020-8291-1} {\bibfield  {journal} {\bibinfo  {journal} {Eur. Phys. J. C}\ }\textbf {\bibinfo {volume} {80}},\ \bibinfo {pages} {708} (\bibinfo {year} {2020})},\ \Eprint {http://arxiv.org/abs/2005.13972} {arXiv:2005.13972 [astro-ph.CO]} \BibitemShut {NoStop}%
\bibitem [{\citenamefont {Feng}\ and\ \citenamefont {Carloni}(2020)}]{Feng:2019dwu}%
  \BibitemOpen
  \bibfield  {author} {\bibinfo {author} {\bibfnamefont {J.~C.}\ \bibnamefont {Feng}}\ and\ \bibinfo {author} {\bibfnamefont {S.}~\bibnamefont {Carloni}},\ }\href {\doibase 10.1103/PhysRevD.101.064002} {\bibfield  {journal} {\bibinfo  {journal} {Phys. Rev. D}\ }\textbf {\bibinfo {volume} {101}},\ \bibinfo {pages} {064002} (\bibinfo {year} {2020})},\ \Eprint {http://arxiv.org/abs/1910.06978} {arXiv:1910.06978 [gr-qc]} \BibitemShut {NoStop}%
\bibitem [{\citenamefont {Feng}\ \emph {et~al.}(2021)\citenamefont {Feng}, \citenamefont {Mukohyama},\ and\ \citenamefont {Carloni}}]{Feng:2020lhp}%
  \BibitemOpen
  \bibfield  {author} {\bibinfo {author} {\bibfnamefont {J.~C.}\ \bibnamefont {Feng}}, \bibinfo {author} {\bibfnamefont {S.}~\bibnamefont {Mukohyama}}, \ and\ \bibinfo {author} {\bibfnamefont {S.}~\bibnamefont {Carloni}},\ }\href {\doibase 10.1103/PhysRevD.103.084055} {\bibfield  {journal} {\bibinfo  {journal} {Phys. Rev. D}\ }\textbf {\bibinfo {volume} {103}},\ \bibinfo {pages} {084055} (\bibinfo {year} {2021})},\ \Eprint {http://arxiv.org/abs/2011.12305} {arXiv:2011.12305 [gr-qc]} \BibitemShut {NoStop}%
\bibitem [{\citenamefont {Danarianto}\ and\ \citenamefont {Sulaksono}(2023{\natexlab{a}})}]{Danarianto:2023ftf}%
  \BibitemOpen
  \bibfield  {author} {\bibinfo {author} {\bibfnamefont {M.~D.}\ \bibnamefont {Danarianto}}\ and\ \bibinfo {author} {\bibfnamefont {A.}~\bibnamefont {Sulaksono}},\ }\href {\doibase 10.1063/5.0178353} {\bibfield  {journal} {\bibinfo  {journal} {AIP Conf. Proc.}\ }\textbf {\bibinfo {volume} {2906}},\ \bibinfo {pages} {030005} (\bibinfo {year} {2023}{\natexlab{a}})}\BibitemShut {NoStop}%
\bibitem [{\citenamefont {Troisi}\ and\ \citenamefont {Carloni}(2025)}]{Troisi:2025ohh}%
  \BibitemOpen
  \bibfield  {author} {\bibinfo {author} {\bibfnamefont {A.}~\bibnamefont {Troisi}}\ and\ \bibinfo {author} {\bibfnamefont {S.}~\bibnamefont {Carloni}},\ }\href {\doibase 10.1016/j.physletb.2025.139426} {\bibfield  {journal} {\bibinfo  {journal} {Phys. Lett. B}\ }\textbf {\bibinfo {volume} {864}},\ \bibinfo {pages} {139426} (\bibinfo {year} {2025})}\BibitemShut {NoStop}%
\bibitem [{\citenamefont {Martins}\ \emph {et~al.}(2021)\citenamefont {Martins}, \citenamefont {Marques}, \citenamefont {Fernandes}, \citenamefont {Oliveira}, \citenamefont {Pinheiro},\ and\ \citenamefont {Rocha}}]{Martins:2021zth}%
  \BibitemOpen
  \bibfield  {author} {\bibinfo {author} {\bibfnamefont {C.~J. A.~P.}\ \bibnamefont {Martins}}, \bibinfo {author} {\bibfnamefont {C.~M.~J.}\ \bibnamefont {Marques}}, \bibinfo {author} {\bibfnamefont {C.~B.~D.}\ \bibnamefont {Fernandes}}, \bibinfo {author} {\bibfnamefont {J.~S. J.~S.}\ \bibnamefont {Oliveira}}, \bibinfo {author} {\bibfnamefont {D.~A.~R.}\ \bibnamefont {Pinheiro}}, \ and\ \bibinfo {author} {\bibfnamefont {B.~A.~R.}\ \bibnamefont {Rocha}},\ }in\ \href {\doibase 10.1142/9789811269776_0070} {\emph {\bibinfo {booktitle} {{16th Marcel Grossmann Meeting on~Recent Developments in Theoretical and Experimental General Relativity, Astrophysics and Relativistic Field Theories}}}}\ (\bibinfo {year} {2021})\ \Eprint {http://arxiv.org/abs/2111.08086} {arXiv:2111.08086 [astro-ph.CO]} \BibitemShut {NoStop}%
\bibitem [{\citenamefont {Fernandes}\ \emph {et~al.}(2021)\citenamefont {Fernandes}, \citenamefont {Martins},\ and\ \citenamefont {Rocha}}]{Fernandes:2020pxf}%
  \BibitemOpen
  \bibfield  {author} {\bibinfo {author} {\bibfnamefont {C.~B.~D.}\ \bibnamefont {Fernandes}}, \bibinfo {author} {\bibfnamefont {C.~J. A.~P.}\ \bibnamefont {Martins}}, \ and\ \bibinfo {author} {\bibfnamefont {B.~A.~R.}\ \bibnamefont {Rocha}},\ }\href {\doibase 10.1016/j.dark.2020.100761} {\bibfield  {journal} {\bibinfo  {journal} {Phys. Dark Univ.}\ }\textbf {\bibinfo {volume} {31}},\ \bibinfo {pages} {100761} (\bibinfo {year} {2021})},\ \Eprint {http://arxiv.org/abs/2012.10513} {arXiv:2012.10513 [astro-ph.CO]} \BibitemShut {NoStop}%
\bibitem [{\citenamefont {Feng}\ \emph {et~al.}(2022)\citenamefont {Feng}, \citenamefont {Mukohyama},\ and\ \citenamefont {Carloni}}]{Feng:2022rga}%
  \BibitemOpen
  \bibfield  {author} {\bibinfo {author} {\bibfnamefont {J.~C.}\ \bibnamefont {Feng}}, \bibinfo {author} {\bibfnamefont {S.}~\bibnamefont {Mukohyama}}, \ and\ \bibinfo {author} {\bibfnamefont {S.}~\bibnamefont {Carloni}},\ }\href {\doibase 10.1103/PhysRevD.105.104036} {\bibfield  {journal} {\bibinfo  {journal} {Phys. Rev. D}\ }\textbf {\bibinfo {volume} {105}},\ \bibinfo {pages} {104036} (\bibinfo {year} {2022})},\ \Eprint {http://arxiv.org/abs/2203.00011} {arXiv:2203.00011 [gr-qc]} \BibitemShut {NoStop}%
\bibitem [{\citenamefont {Danarianto}\ and\ \citenamefont {Sulaksono}(2023{\natexlab{b}})}]{Danarianto:2023rff}%
  \BibitemOpen
  \bibfield  {author} {\bibinfo {author} {\bibfnamefont {M.~D.}\ \bibnamefont {Danarianto}}\ and\ \bibinfo {author} {\bibfnamefont {A.}~\bibnamefont {Sulaksono}},\ }\href {\doibase 10.1140/epjc/s10052-023-11644-2} {\bibfield  {journal} {\bibinfo  {journal} {Eur. Phys. J. C}\ }\textbf {\bibinfo {volume} {83}},\ \bibinfo {pages} {463} (\bibinfo {year} {2023}{\natexlab{b}})}\BibitemShut {NoStop}%
\bibitem [{\citenamefont {Chandrasekhar}(1972)}]{chandrasekhar:1972lre}%
  \BibitemOpen
  \bibfield  {author} {\bibinfo {author} {\bibfnamefont {S.}~\bibnamefont {Chandrasekhar}},\ }in\ \href@noop {} {\emph {\bibinfo {booktitle} {General relativity: papers in honour of J. L. Synge}}},\ \bibinfo {editor} {edited by\ \bibinfo {editor} {\bibfnamefont {L.}~\bibnamefont {O'Raifeartaigh}}}\ (\bibinfo  {publisher} {Clarendon Press},\ \bibinfo {address} {Oxford},\ \bibinfo {year} {1972})\ pp.\ \bibinfo {pages} {185--199}\BibitemShut {NoStop}%
\bibitem [{\citenamefont {Chavanis}(2002)}]{chavanis:2002iis}%
  \BibitemOpen
  \bibfield  {author} {\bibinfo {author} {\bibfnamefont {P.-H.}\ \bibnamefont {Chavanis}},\ }\href@noop {} {\bibfield  {journal} {\bibinfo  {journal} {Astronomy \& Astrophysics}\ }\textbf {\bibinfo {volume} {381}},\ \bibinfo {pages} {709} (\bibinfo {year} {2002})}\BibitemShut {NoStop}%
\bibitem [{\citenamefont {Weinberg}(1989)}]{Weinberg:1988cp}%
  \BibitemOpen
  \bibfield  {author} {\bibinfo {author} {\bibfnamefont {S.}~\bibnamefont {Weinberg}},\ }\href {\doibase 10.1103/RevModPhys.61.1} {\bibfield  {journal} {\bibinfo  {journal} {Rev. Mod. Phys.}\ }\textbf {\bibinfo {volume} {61}},\ \bibinfo {pages} {1} (\bibinfo {year} {1989})}\BibitemShut {NoStop}%
\bibitem [{\citenamefont {Chandrasekhar}(1939)}]{Chandrasekhar:1939}%
  \BibitemOpen
  \bibfield  {author} {\bibinfo {author} {\bibfnamefont {S.}~\bibnamefont {Chandrasekhar}},\ }\href@noop {} {\emph {\bibinfo {title} {{An introduction to the study of stellar structure}}}}\ (\bibinfo  {publisher} {Chicago Univ. Pr.},\ \bibinfo {address} {Chicago, USA},\ \bibinfo {year} {1939})\BibitemShut {NoStop}%
\bibitem [{\citenamefont {Pasechnik}\ and\ \citenamefont {\v{S}umbera}(2017)}]{Pasechnik:2016wkt}%
  \BibitemOpen
  \bibfield  {author} {\bibinfo {author} {\bibfnamefont {R.}~\bibnamefont {Pasechnik}}\ and\ \bibinfo {author} {\bibfnamefont {M.}~\bibnamefont {\v{S}umbera}},\ }\href {\doibase 10.3390/universe3010007} {\bibfield  {journal} {\bibinfo  {journal} {Universe}\ }\textbf {\bibinfo {volume} {3}},\ \bibinfo {pages} {7} (\bibinfo {year} {2017})},\ \Eprint {http://arxiv.org/abs/1611.01533} {arXiv:1611.01533 [hep-ph]} \BibitemShut {NoStop}%
\bibitem [{\citenamefont {Carr}\ \emph {et~al.}(2021)\citenamefont {Carr}, \citenamefont {Kohri}, \citenamefont {Sendouda},\ and\ \citenamefont {Yokoyama}}]{Carr:2020gox}%
  \BibitemOpen
  \bibfield  {author} {\bibinfo {author} {\bibfnamefont {B.}~\bibnamefont {Carr}}, \bibinfo {author} {\bibfnamefont {K.}~\bibnamefont {Kohri}}, \bibinfo {author} {\bibfnamefont {Y.}~\bibnamefont {Sendouda}}, \ and\ \bibinfo {author} {\bibfnamefont {J.}~\bibnamefont {Yokoyama}},\ }\href {\doibase 10.1088/1361-6633/ac1e31} {\bibfield  {journal} {\bibinfo  {journal} {Rept. Prog. Phys.}\ }\textbf {\bibinfo {volume} {84}},\ \bibinfo {pages} {116902} (\bibinfo {year} {2021})},\ \Eprint {http://arxiv.org/abs/2002.12778} {arXiv:2002.12778 [astro-ph.CO]} \BibitemShut {NoStop}%
\bibitem [{\citenamefont {Hawking}(1989)}]{Hawking:1987bn}%
  \BibitemOpen
  \bibfield  {author} {\bibinfo {author} {\bibfnamefont {S.~W.}\ \bibnamefont {Hawking}},\ }\href {\doibase 10.1016/0370-2693(89)90206-2} {\bibfield  {journal} {\bibinfo  {journal} {Phys. Lett. B}\ }\textbf {\bibinfo {volume} {231}},\ \bibinfo {pages} {237} (\bibinfo {year} {1989})}\BibitemShut {NoStop}%
\bibitem [{\citenamefont {Abrahams}\ and\ \citenamefont {Evans}(1992)}]{Abrahams:1992ib}%
  \BibitemOpen
  \bibfield  {author} {\bibinfo {author} {\bibfnamefont {A.~M.}\ \bibnamefont {Abrahams}}\ and\ \bibinfo {author} {\bibfnamefont {C.~R.}\ \bibnamefont {Evans}},\ }\href {\doibase 10.1103/PhysRevD.46.R4117} {\bibfield  {journal} {\bibinfo  {journal} {Phys. Rev. D}\ }\textbf {\bibinfo {volume} {46}},\ \bibinfo {pages} {R4117} (\bibinfo {year} {1992})}\BibitemShut {NoStop}%
\bibitem [{\citenamefont {Alcubierre}\ \emph {et~al.}(2000)\citenamefont {Alcubierre}, \citenamefont {Allen}, \citenamefont {Bruegmann}, \citenamefont {Lanfermann}, \citenamefont {Seidel}, \citenamefont {Suen},\ and\ \citenamefont {Tobias}}]{Alcubierre:1999ex}%
  \BibitemOpen
  \bibfield  {author} {\bibinfo {author} {\bibfnamefont {M.}~\bibnamefont {Alcubierre}}, \bibinfo {author} {\bibfnamefont {G.}~\bibnamefont {Allen}}, \bibinfo {author} {\bibfnamefont {B.}~\bibnamefont {Bruegmann}}, \bibinfo {author} {\bibfnamefont {G.}~\bibnamefont {Lanfermann}}, \bibinfo {author} {\bibfnamefont {E.}~\bibnamefont {Seidel}}, \bibinfo {author} {\bibfnamefont {W.-M.}\ \bibnamefont {Suen}}, \ and\ \bibinfo {author} {\bibfnamefont {M.}~\bibnamefont {Tobias}},\ }\href {\doibase 10.1103/PhysRevD.61.041501} {\bibfield  {journal} {\bibinfo  {journal} {Phys. Rev. D}\ }\textbf {\bibinfo {volume} {61}},\ \bibinfo {pages} {041501} (\bibinfo {year} {2000})},\ \Eprint {http://arxiv.org/abs/gr-qc/9904013} {arXiv:gr-qc/9904013} \BibitemShut {NoStop}%
\bibitem [{\citenamefont {Adler}\ \emph {et~al.}(2001{\natexlab{a}})\citenamefont {Adler}, \citenamefont {Chen},\ and\ \citenamefont {Santiago}}]{adler2001generalized}%
  \BibitemOpen
  \bibfield  {author} {\bibinfo {author} {\bibfnamefont {R.~J.}\ \bibnamefont {Adler}}, \bibinfo {author} {\bibfnamefont {P.}~\bibnamefont {Chen}}, \ and\ \bibinfo {author} {\bibfnamefont {D.~I.}\ \bibnamefont {Santiago}},\ }\href@noop {} {\bibfield  {journal} {\bibinfo  {journal} {General Relativity and Gravitation}\ }\textbf {\bibinfo {volume} {33}},\ \bibinfo {pages} {2101} (\bibinfo {year} {2001}{\natexlab{a}})}\BibitemShut {NoStop}%
\bibitem [{\citenamefont {Chen}\ and\ \citenamefont {Adler}(2003{\natexlab{a}})}]{chen2003black}%
  \BibitemOpen
  \bibfield  {author} {\bibinfo {author} {\bibfnamefont {P.}~\bibnamefont {Chen}}\ and\ \bibinfo {author} {\bibfnamefont {R.~J.}\ \bibnamefont {Adler}},\ }\href@noop {} {\bibfield  {journal} {\bibinfo  {journal} {Nuclear Physics B-Proceedings Supplements}\ }\textbf {\bibinfo {volume} {124}},\ \bibinfo {pages} {103} (\bibinfo {year} {2003}{\natexlab{a}})}\BibitemShut {NoStop}%
\bibitem [{\citenamefont {Kavanagh}(2019)}]{Kavanagh:2019}%
  \BibitemOpen
  \bibfield  {author} {\bibinfo {author} {\bibfnamefont {B.~J.}\ \bibnamefont {Kavanagh}},\ }\href {\doibase 10.5281/zenodo.3538998} {\enquote {\bibinfo {title} {bradkav/pbhbounds: Release version},}\ } (\bibinfo {year} {2019})\BibitemShut {NoStop}%
\bibitem [{\citenamefont {Rackauckas}\ and\ \citenamefont {Nie}(2017)}]{SciMLDiffEq:2017}%
  \BibitemOpen
  \bibfield  {author} {\bibinfo {author} {\bibfnamefont {C.}~\bibnamefont {Rackauckas}}\ and\ \bibinfo {author} {\bibfnamefont {Q.}~\bibnamefont {Nie}},\ }\href {\doibase 10.5334/jors.151} {\bibfield  {journal} {\bibinfo  {journal} {Journal of Open Research Software}\ }\textbf {\bibinfo {volume} {5}} (\bibinfo {year} {2017}),\ 10.5334/jors.151},\ \bibinfo {note} {{Accessed 2024-07-03}}\BibitemShut {NoStop}%
\bibitem [{\citenamefont {Pretel}(2020)}]{Pretel:2020xuo}%
  \BibitemOpen
  \bibfield  {author} {\bibinfo {author} {\bibfnamefont {J.~M.~Z.}\ \bibnamefont {Pretel}},\ }\href {\doibase 10.1140/epjc/s10052-020-8301-3} {\bibfield  {journal} {\bibinfo  {journal} {Eur. Phys. J. C}\ }\textbf {\bibinfo {volume} {80}},\ \bibinfo {pages} {726} (\bibinfo {year} {2020})},\ \Eprint {http://arxiv.org/abs/2008.05331} {arXiv:2008.05331 [gr-qc]} \BibitemShut {NoStop}%
\bibitem [{\citenamefont {Shapiro}\ and\ \citenamefont {Teukolsky}(1983)}]{Shapiro:1983du}%
  \BibitemOpen
  \bibfield  {author} {\bibinfo {author} {\bibfnamefont {S.~L.}\ \bibnamefont {Shapiro}}\ and\ \bibinfo {author} {\bibfnamefont {S.~A.}\ \bibnamefont {Teukolsky}},\ }\href {\doibase 10.1002/9783527617661} {\emph {\bibinfo {title} {{Black holes, white dwarfs, and neutron stars: The physics of compact objects}}}}\ (\bibinfo {year} {1983})\BibitemShut {NoStop}%
\bibitem [{\citenamefont {Chandrasekhar}(1964{\natexlab{a}})}]{Chandrasekhar:1964zz}%
  \BibitemOpen
  \bibfield  {author} {\bibinfo {author} {\bibfnamefont {S.}~\bibnamefont {Chandrasekhar}},\ }\href {\doibase 10.1086/147938} {\bibfield  {journal} {\bibinfo  {journal} {Astrophys. J.}\ }\textbf {\bibinfo {volume} {140}},\ \bibinfo {pages} {417} (\bibinfo {year} {1964}{\natexlab{a}})},\ \bibinfo {note} {[Erratum: Astrophys.J. 140, 1342 (1964)]}\BibitemShut {NoStop}%
\bibitem [{\citenamefont {Chandrasekhar}(1964{\natexlab{b}})}]{Chandrasekhar:1964zza}%
  \BibitemOpen
  \bibfield  {author} {\bibinfo {author} {\bibfnamefont {S.}~\bibnamefont {Chandrasekhar}},\ }\href {\doibase 10.1103/PhysRevLett.12.114} {\bibfield  {journal} {\bibinfo  {journal} {Phys. Rev. Lett.}\ }\textbf {\bibinfo {volume} {12}},\ \bibinfo {pages} {114} (\bibinfo {year} {1964}{\natexlab{b}})}\BibitemShut {NoStop}%
\bibitem [{\citenamefont {Chanmugam}(1977)}]{chanmugam1977radial}%
  \BibitemOpen
  \bibfield  {author} {\bibinfo {author} {\bibfnamefont {G.}~\bibnamefont {Chanmugam}},\ }\href@noop {} {\bibfield  {journal} {\bibinfo  {journal} {Astrophys. J.}\ }\textbf {\bibinfo {volume} {217}},\ \bibinfo {pages} {799} (\bibinfo {year} {1977})}\BibitemShut {NoStop}%
\bibitem [{\citenamefont {Luz}\ and\ \citenamefont {Carloni}(2024{\natexlab{a}})}]{Luz:2024lgi}%
  \BibitemOpen
  \bibfield  {author} {\bibinfo {author} {\bibfnamefont {P.}~\bibnamefont {Luz}}\ and\ \bibinfo {author} {\bibfnamefont {S.}~\bibnamefont {Carloni}},\ }\href {\doibase 10.1103/PhysRevD.110.084055} {\bibfield  {journal} {\bibinfo  {journal} {Phys. Rev. D}\ }\textbf {\bibinfo {volume} {110}},\ \bibinfo {pages} {084055} (\bibinfo {year} {2024}{\natexlab{a}})},\ \Eprint {http://arxiv.org/abs/2405.10359} {arXiv:2405.10359 [gr-qc]} \BibitemShut {NoStop}%
\bibitem [{\citenamefont {Luz}\ and\ \citenamefont {Carloni}(2024{\natexlab{b}})}]{Luz:2024xnd}%
  \BibitemOpen
  \bibfield  {author} {\bibinfo {author} {\bibfnamefont {P.}~\bibnamefont {Luz}}\ and\ \bibinfo {author} {\bibfnamefont {S.}~\bibnamefont {Carloni}},\ }\href {\doibase 10.1103/PhysRevD.110.084054} {\bibfield  {journal} {\bibinfo  {journal} {Phys. Rev. D}\ }\textbf {\bibinfo {volume} {110}},\ \bibinfo {pages} {084054} (\bibinfo {year} {2024}{\natexlab{b}})},\ \Eprint {http://arxiv.org/abs/2405.06740} {arXiv:2405.06740 [gr-qc]} \BibitemShut {NoStop}%
\bibitem [{\citenamefont {Luz}\ and\ \citenamefont {Carloni}(2024{\natexlab{c}})}]{Luz:2024yjm}%
  \BibitemOpen
  \bibfield  {author} {\bibinfo {author} {\bibfnamefont {P.}~\bibnamefont {Luz}}\ and\ \bibinfo {author} {\bibfnamefont {S.}~\bibnamefont {Carloni}},\ }\href {\doibase 10.1088/1361-6382/ad8a14} {\bibfield  {journal} {\bibinfo  {journal} {Class. Quant. Grav.}\ }\textbf {\bibinfo {volume} {41}},\ \bibinfo {pages} {235012} (\bibinfo {year} {2024}{\natexlab{c}})},\ \Eprint {http://arxiv.org/abs/2405.05321} {arXiv:2405.05321 [gr-qc]} \BibitemShut {NoStop}%
\bibitem [{\citenamefont {Santiago}\ \emph {et~al.}(2025)\citenamefont {Santiago}, \citenamefont {Feng}, \citenamefont {Schuster},\ and\ \citenamefont {Visser}}]{Santiago:2025rzb}%
  \BibitemOpen
  \bibfield  {author} {\bibinfo {author} {\bibfnamefont {J.}~\bibnamefont {Santiago}}, \bibinfo {author} {\bibfnamefont {J.}~\bibnamefont {Feng}}, \bibinfo {author} {\bibfnamefont {S.}~\bibnamefont {Schuster}}, \ and\ \bibinfo {author} {\bibfnamefont {M.}~\bibnamefont {Visser}},\ }\href@noop {} {\  (\bibinfo {year} {2025})},\ \Eprint {http://arxiv.org/abs/2503.20696} {arXiv:2503.20696 [gr-qc]} \BibitemShut {NoStop}%
\bibitem [{\citenamefont {Baker}\ \emph {et~al.}(2025)\citenamefont {Baker}, \citenamefont {Iguaz~Juan}, \citenamefont {Symons},\ and\ \citenamefont {Thamm}}]{Baker:2025zxm}%
  \BibitemOpen
  \bibfield  {author} {\bibinfo {author} {\bibfnamefont {M.~J.}\ \bibnamefont {Baker}}, \bibinfo {author} {\bibfnamefont {J.}~\bibnamefont {Iguaz~Juan}}, \bibinfo {author} {\bibfnamefont {A.}~\bibnamefont {Symons}}, \ and\ \bibinfo {author} {\bibfnamefont {A.}~\bibnamefont {Thamm}},\ }\href@noop {} {\  (\bibinfo {year} {2025})},\ \Eprint {http://arxiv.org/abs/2503.10755} {arXiv:2503.10755 [hep-ph]} \BibitemShut {NoStop}%
\bibitem [{\citenamefont {Davies}\ \emph {et~al.}(2024)\citenamefont {Davies}, \citenamefont {Easson},\ and\ \citenamefont {Levin}}]{Davies:2024ysj}%
  \BibitemOpen
  \bibfield  {author} {\bibinfo {author} {\bibfnamefont {P.~C.~W.}\ \bibnamefont {Davies}}, \bibinfo {author} {\bibfnamefont {D.~A.}\ \bibnamefont {Easson}}, \ and\ \bibinfo {author} {\bibfnamefont {P.~B.}\ \bibnamefont {Levin}},\ }\href@noop {} {\  (\bibinfo {year} {2024})},\ \Eprint {http://arxiv.org/abs/2410.21577} {arXiv:2410.21577 [hep-th]} \BibitemShut {NoStop}%
\bibitem [{\citenamefont {Carr}\ and\ \citenamefont {Kuhnel}(2022)}]{Carr:2021bzv}%
  \BibitemOpen
  \bibfield  {author} {\bibinfo {author} {\bibfnamefont {B.}~\bibnamefont {Carr}}\ and\ \bibinfo {author} {\bibfnamefont {F.}~\bibnamefont {Kuhnel}},\ }\href {\doibase 10.21468/SciPostPhysLectNotes.48} {\bibfield  {journal} {\bibinfo  {journal} {SciPost Phys. Lect. Notes}\ }\textbf {\bibinfo {volume} {48}},\ \bibinfo {pages} {1} (\bibinfo {year} {2022})},\ \Eprint {http://arxiv.org/abs/2110.02821} {arXiv:2110.02821 [astro-ph.CO]} \BibitemShut {NoStop}%
\bibitem [{\citenamefont {Bai}\ and\ \citenamefont {Orlofsky}(2020)}]{Bai:2019zcd}%
  \BibitemOpen
  \bibfield  {author} {\bibinfo {author} {\bibfnamefont {Y.}~\bibnamefont {Bai}}\ and\ \bibinfo {author} {\bibfnamefont {N.}~\bibnamefont {Orlofsky}},\ }\href {\doibase 10.1103/PhysRevD.101.055006} {\bibfield  {journal} {\bibinfo  {journal} {Phys. Rev. D}\ }\textbf {\bibinfo {volume} {101}},\ \bibinfo {pages} {055006} (\bibinfo {year} {2020})},\ \Eprint {http://arxiv.org/abs/1906.04858} {arXiv:1906.04858 [hep-ph]} \BibitemShut {NoStop}%
\bibitem [{\citenamefont {Chen}\ and\ \citenamefont {Adler}(2003{\natexlab{b}})}]{Chen:2002tu}%
  \BibitemOpen
  \bibfield  {author} {\bibinfo {author} {\bibfnamefont {P.}~\bibnamefont {Chen}}\ and\ \bibinfo {author} {\bibfnamefont {R.~J.}\ \bibnamefont {Adler}},\ }\href {\doibase 10.1016/S0920-5632(03)02088-7} {\bibfield  {journal} {\bibinfo  {journal} {Nucl. Phys. B Proc. Suppl.}\ }\textbf {\bibinfo {volume} {124}},\ \bibinfo {pages} {103} (\bibinfo {year} {2003}{\natexlab{b}})},\ \Eprint {http://arxiv.org/abs/gr-qc/0205106} {arXiv:gr-qc/0205106} \BibitemShut {NoStop}%
\bibitem [{\citenamefont {Adler}\ \emph {et~al.}(2001{\natexlab{b}})\citenamefont {Adler}, \citenamefont {Chen},\ and\ \citenamefont {Santiago}}]{Adler:2001vs}%
  \BibitemOpen
  \bibfield  {author} {\bibinfo {author} {\bibfnamefont {R.~J.}\ \bibnamefont {Adler}}, \bibinfo {author} {\bibfnamefont {P.}~\bibnamefont {Chen}}, \ and\ \bibinfo {author} {\bibfnamefont {D.~I.}\ \bibnamefont {Santiago}},\ }\href {\doibase 10.1023/A:1015281430411} {\bibfield  {journal} {\bibinfo  {journal} {Gen. Rel. Grav.}\ }\textbf {\bibinfo {volume} {33}},\ \bibinfo {pages} {2101} (\bibinfo {year} {2001}{\natexlab{b}})},\ \Eprint {http://arxiv.org/abs/gr-qc/0106080} {arXiv:gr-qc/0106080} \BibitemShut {NoStop}%
\bibitem [{\citenamefont {Barrow}\ \emph {et~al.}(1992)\citenamefont {Barrow}, \citenamefont {Copeland},\ and\ \citenamefont {Liddle}}]{Barrow:1992hq}%
  \BibitemOpen
  \bibfield  {author} {\bibinfo {author} {\bibfnamefont {J.~D.}\ \bibnamefont {Barrow}}, \bibinfo {author} {\bibfnamefont {E.~J.}\ \bibnamefont {Copeland}}, \ and\ \bibinfo {author} {\bibfnamefont {A.~R.}\ \bibnamefont {Liddle}},\ }\href {\doibase 10.1103/PhysRevD.46.645} {\bibfield  {journal} {\bibinfo  {journal} {Phys. Rev. D}\ }\textbf {\bibinfo {volume} {46}},\ \bibinfo {pages} {645} (\bibinfo {year} {1992})}\BibitemShut {NoStop}%
\bibitem [{\citenamefont {Barrow}(1992)}]{Barrow:1992ay}%
  \BibitemOpen
  \bibfield  {author} {\bibinfo {author} {\bibfnamefont {J.~D.}\ \bibnamefont {Barrow}},\ }\href {\doibase 10.1103/PhysRevD.47.1730} {\bibfield  {journal} {\bibinfo  {journal} {Phys. Rev. D}\ }\textbf {\bibinfo {volume} {46}},\ \bibinfo {pages} {R3227} (\bibinfo {year} {1992})},\ \bibinfo {note} {[Erratum: Phys.Rev.D 47, 1730 (1993)]}\BibitemShut {NoStop}%
\bibitem [{\citenamefont {Aharonov}\ \emph {et~al.}(1987)\citenamefont {Aharonov}, \citenamefont {Casher},\ and\ \citenamefont {Nussinov}}]{Aharonov:1987tp}%
  \BibitemOpen
  \bibfield  {author} {\bibinfo {author} {\bibfnamefont {Y.}~\bibnamefont {Aharonov}}, \bibinfo {author} {\bibfnamefont {A.}~\bibnamefont {Casher}}, \ and\ \bibinfo {author} {\bibfnamefont {S.}~\bibnamefont {Nussinov}},\ }\href {\doibase 10.1016/0370-2693(87)91320-7} {\bibfield  {journal} {\bibinfo  {journal} {Phys. Lett. B}\ }\textbf {\bibinfo {volume} {191}},\ \bibinfo {pages} {51} (\bibinfo {year} {1987})}\BibitemShut {NoStop}%
\bibitem [{\citenamefont {Dvali}\ \emph {et~al.}(2025)\citenamefont {Dvali}, \citenamefont {Zantedeschi},\ and\ \citenamefont {Zell}}]{Dvali:2025ktz}%
  \BibitemOpen
  \bibfield  {author} {\bibinfo {author} {\bibfnamefont {G.}~\bibnamefont {Dvali}}, \bibinfo {author} {\bibfnamefont {M.}~\bibnamefont {Zantedeschi}}, \ and\ \bibinfo {author} {\bibfnamefont {S.}~\bibnamefont {Zell}},\ }\href@noop {} {\  (\bibinfo {year} {2025})},\ \Eprint {http://arxiv.org/abs/2503.21740} {arXiv:2503.21740 [hep-ph]} \BibitemShut {NoStop}%
\bibitem [{\citenamefont {Thoss}\ \emph {et~al.}(2024)\citenamefont {Thoss}, \citenamefont {Burkert},\ and\ \citenamefont {Kohri}}]{Thoss:2024hsr}%
  \BibitemOpen
  \bibfield  {author} {\bibinfo {author} {\bibfnamefont {V.}~\bibnamefont {Thoss}}, \bibinfo {author} {\bibfnamefont {A.}~\bibnamefont {Burkert}}, \ and\ \bibinfo {author} {\bibfnamefont {K.}~\bibnamefont {Kohri}},\ }\href {\doibase 10.1093/mnras/stae1098} {\bibfield  {journal} {\bibinfo  {journal} {Mon. Not. Roy. Astron. Soc.}\ }\textbf {\bibinfo {volume} {532}},\ \bibinfo {pages} {451} (\bibinfo {year} {2024})},\ \Eprint {http://arxiv.org/abs/2402.17823} {arXiv:2402.17823 [astro-ph.CO]} \BibitemShut {NoStop}%
\bibitem [{\citenamefont {Dvali}\ \emph {et~al.}(2024)\citenamefont {Dvali}, \citenamefont {Valbuena-Berm\'udez},\ and\ \citenamefont {Zantedeschi}}]{Dvali:2024hsb}%
  \BibitemOpen
  \bibfield  {author} {\bibinfo {author} {\bibfnamefont {G.}~\bibnamefont {Dvali}}, \bibinfo {author} {\bibfnamefont {J.~S.}\ \bibnamefont {Valbuena-Berm\'udez}}, \ and\ \bibinfo {author} {\bibfnamefont {M.}~\bibnamefont {Zantedeschi}},\ }\href {\doibase 10.1103/PhysRevD.110.056029} {\bibfield  {journal} {\bibinfo  {journal} {Phys. Rev. D}\ }\textbf {\bibinfo {volume} {110}},\ \bibinfo {pages} {056029} (\bibinfo {year} {2024})},\ \Eprint {http://arxiv.org/abs/2405.13117} {arXiv:2405.13117 [hep-th]} \BibitemShut {NoStop}%
\bibitem [{\citenamefont {Alexandre}\ \emph {et~al.}(2024)\citenamefont {Alexandre}, \citenamefont {Dvali},\ and\ \citenamefont {Koutsangelas}}]{Alexandre:2024nuo}%
  \BibitemOpen
  \bibfield  {author} {\bibinfo {author} {\bibfnamefont {A.}~\bibnamefont {Alexandre}}, \bibinfo {author} {\bibfnamefont {G.}~\bibnamefont {Dvali}}, \ and\ \bibinfo {author} {\bibfnamefont {E.}~\bibnamefont {Koutsangelas}},\ }\href {\doibase 10.1103/PhysRevD.110.036004} {\bibfield  {journal} {\bibinfo  {journal} {Phys. Rev. D}\ }\textbf {\bibinfo {volume} {110}},\ \bibinfo {pages} {036004} (\bibinfo {year} {2024})},\ \Eprint {http://arxiv.org/abs/2402.14069} {arXiv:2402.14069 [hep-ph]} \BibitemShut {NoStop}%
\bibitem [{\citenamefont {Dvali}\ \emph {et~al.}(2020)\citenamefont {Dvali}, \citenamefont {Eisemann}, \citenamefont {Michel},\ and\ \citenamefont {Zell}}]{Dvali:2020wft}%
  \BibitemOpen
  \bibfield  {author} {\bibinfo {author} {\bibfnamefont {G.}~\bibnamefont {Dvali}}, \bibinfo {author} {\bibfnamefont {L.}~\bibnamefont {Eisemann}}, \bibinfo {author} {\bibfnamefont {M.}~\bibnamefont {Michel}}, \ and\ \bibinfo {author} {\bibfnamefont {S.}~\bibnamefont {Zell}},\ }\href {\doibase 10.1103/PhysRevD.102.103523} {\bibfield  {journal} {\bibinfo  {journal} {Phys. Rev. D}\ }\textbf {\bibinfo {volume} {102}},\ \bibinfo {pages} {103523} (\bibinfo {year} {2020})},\ \Eprint {http://arxiv.org/abs/2006.00011} {arXiv:2006.00011 [hep-th]} \BibitemShut {NoStop}%
\bibitem [{\citenamefont {Dvali}(2018)}]{Dvali:2018xpy}%
  \BibitemOpen
  \bibfield  {author} {\bibinfo {author} {\bibfnamefont {G.}~\bibnamefont {Dvali}},\ }\href@noop {} {\  (\bibinfo {year} {2018})},\ \Eprint {http://arxiv.org/abs/1810.02336} {arXiv:1810.02336 [hep-th]} \BibitemShut {NoStop}%
\bibitem [{\citenamefont {Alcock}\ \emph {et~al.}(2000)\citenamefont {Alcock} \emph {et~al.}}]{MACHO:2000qbb}%
  \BibitemOpen
  \bibfield  {author} {\bibinfo {author} {\bibfnamefont {C.}~\bibnamefont {Alcock}} \emph {et~al.} (\bibinfo {collaboration} {MACHO}),\ }\href {\doibase 10.1086/309512} {\bibfield  {journal} {\bibinfo  {journal} {Astrophys. J.}\ }\textbf {\bibinfo {volume} {542}},\ \bibinfo {pages} {281} (\bibinfo {year} {2000})},\ \Eprint {http://arxiv.org/abs/astro-ph/0001272} {arXiv:astro-ph/0001272} \BibitemShut {NoStop}%
\bibitem [{\citenamefont {Basak}\ \emph {et~al.}(2022)\citenamefont {Basak}, \citenamefont {Ganguly}, \citenamefont {Haris}, \citenamefont {Kapadia}, \citenamefont {Mehta},\ and\ \citenamefont {Ajith}}]{Basak:2021ten}%
  \BibitemOpen
  \bibfield  {author} {\bibinfo {author} {\bibfnamefont {S.}~\bibnamefont {Basak}}, \bibinfo {author} {\bibfnamefont {A.}~\bibnamefont {Ganguly}}, \bibinfo {author} {\bibfnamefont {K.}~\bibnamefont {Haris}}, \bibinfo {author} {\bibfnamefont {S.}~\bibnamefont {Kapadia}}, \bibinfo {author} {\bibfnamefont {A.~K.}\ \bibnamefont {Mehta}}, \ and\ \bibinfo {author} {\bibfnamefont {P.}~\bibnamefont {Ajith}},\ }\href {\doibase 10.3847/2041-8213/ac4dfa} {\bibfield  {journal} {\bibinfo  {journal} {Astrophys. J.}\ }\textbf {\bibinfo {volume} {926}},\ \bibinfo {pages} {L28} (\bibinfo {year} {2022})},\ \Eprint {http://arxiv.org/abs/2109.06456} {arXiv:2109.06456 [gr-qc]} \BibitemShut {NoStop}%
\bibitem [{\citenamefont {Sammons}\ \emph {et~al.}(2020)\citenamefont {Sammons}, \citenamefont {Macquart}, \citenamefont {Ekers}, \citenamefont {Shannon}, \citenamefont {Cho}, \citenamefont {Prochaska}, \citenamefont {Deller},\ and\ \citenamefont {Day}}]{Sammons:2020kyk}%
  \BibitemOpen
  \bibfield  {author} {\bibinfo {author} {\bibfnamefont {M.~W.}\ \bibnamefont {Sammons}}, \bibinfo {author} {\bibfnamefont {J.-P.}\ \bibnamefont {Macquart}}, \bibinfo {author} {\bibfnamefont {R.~D.}\ \bibnamefont {Ekers}}, \bibinfo {author} {\bibfnamefont {R.~M.}\ \bibnamefont {Shannon}}, \bibinfo {author} {\bibfnamefont {H.}~\bibnamefont {Cho}}, \bibinfo {author} {\bibfnamefont {J.~X.}\ \bibnamefont {Prochaska}}, \bibinfo {author} {\bibfnamefont {A.~T.}\ \bibnamefont {Deller}}, \ and\ \bibinfo {author} {\bibfnamefont {C.~K.}\ \bibnamefont {Day}},\ }\href {\doibase 10.3847/1538-4357/aba7bb} {\bibfield  {journal} {\bibinfo  {journal} {Astrophys. J.}\ }\textbf {\bibinfo {volume} {900}},\ \bibinfo {pages} {122} (\bibinfo {year} {2020})},\ \Eprint {http://arxiv.org/abs/2002.12533} {arXiv:2002.12533 [astro-ph.CO]} \BibitemShut {NoStop}%
\bibitem [{\citenamefont {Shah}\ \emph {et~al.}(2024)\citenamefont {Shah} \emph {et~al.}}]{DES:2024ffp}%
  \BibitemOpen
  \bibfield  {author} {\bibinfo {author} {\bibfnamefont {P.}~\bibnamefont {Shah}} \emph {et~al.} (\bibinfo {collaboration} {DES}),\ }\href {\doibase 10.1093/mnras/stae2614} {\bibfield  {journal} {\bibinfo  {journal} {Mon. Not. Roy. Astron. Soc.}\ }\textbf {\bibinfo {volume} {536}},\ \bibinfo {pages} {946} (\bibinfo {year} {2024})},\ \Eprint {http://arxiv.org/abs/2410.07956} {arXiv:2410.07956 [astro-ph.CO]} \BibitemShut {NoStop}%
\bibitem [{\citenamefont {Brandt}(2016)}]{Brandt:2016aco}%
  \BibitemOpen
  \bibfield  {author} {\bibinfo {author} {\bibfnamefont {T.~D.}\ \bibnamefont {Brandt}},\ }\href {\doibase 10.3847/2041-8205/824/2/L31} {\bibfield  {journal} {\bibinfo  {journal} {Astrophys. J. Lett.}\ }\textbf {\bibinfo {volume} {824}},\ \bibinfo {pages} {L31} (\bibinfo {year} {2016})},\ \Eprint {http://arxiv.org/abs/1605.03665} {arXiv:1605.03665 [astro-ph.GA]} \BibitemShut {NoStop}%
\bibitem [{\citenamefont {Liebling}\ and\ \citenamefont {Palenzuela}(2023)}]{Liebling:2012fv}%
  \BibitemOpen
  \bibfield  {author} {\bibinfo {author} {\bibfnamefont {S.~L.}\ \bibnamefont {Liebling}}\ and\ \bibinfo {author} {\bibfnamefont {C.}~\bibnamefont {Palenzuela}},\ }\href {\doibase 10.1007/s41114-023-00043-4} {\bibfield  {journal} {\bibinfo  {journal} {Living Rev. Rel.}\ }\textbf {\bibinfo {volume} {26}},\ \bibinfo {pages} {1} (\bibinfo {year} {2023})},\ \Eprint {http://arxiv.org/abs/1202.5809} {arXiv:1202.5809 [gr-qc]} \BibitemShut {NoStop}%
\bibitem [{\citenamefont {Visinelli}(2021)}]{Visinelli:2021uve}%
  \BibitemOpen
  \bibfield  {author} {\bibinfo {author} {\bibfnamefont {L.}~\bibnamefont {Visinelli}},\ }\href {\doibase 10.1142/S0218271821300068} {\bibfield  {journal} {\bibinfo  {journal} {Int. J. Mod. Phys. D}\ }\textbf {\bibinfo {volume} {30}},\ \bibinfo {pages} {2130006} (\bibinfo {year} {2021})},\ \Eprint {http://arxiv.org/abs/2109.05481} {arXiv:2109.05481 [gr-qc]} \BibitemShut {NoStop}%
\bibitem [{\citenamefont {Kaup}(1968)}]{Kaup:1968}%
  \BibitemOpen
  \bibfield  {author} {\bibinfo {author} {\bibfnamefont {D.~J.}\ \bibnamefont {Kaup}},\ }\href {\doibase 10.1103/PhysRev.172.1331} {\bibfield  {journal} {\bibinfo  {journal} {Phys. Rev.}\ }\textbf {\bibinfo {volume} {172}},\ \bibinfo {pages} {1331} (\bibinfo {year} {1968})}\BibitemShut {NoStop}%
\bibitem [{\citenamefont {Bagui}\ \emph {et~al.}(2025)\citenamefont {Bagui} \emph {et~al.}}]{LISACosmologyWorkingGroup:2023njw}%
  \BibitemOpen
  \bibfield  {author} {\bibinfo {author} {\bibfnamefont {E.}~\bibnamefont {Bagui}} \emph {et~al.} (\bibinfo {collaboration} {LISA Cosmology Working Group}),\ }\href {\doibase 10.1007/s41114-024-00053-w} {\bibfield  {journal} {\bibinfo  {journal} {Living Rev. Rel.}\ }\textbf {\bibinfo {volume} {28}},\ \bibinfo {pages} {1} (\bibinfo {year} {2025})},\ \Eprint {http://arxiv.org/abs/2310.19857} {arXiv:2310.19857 [astro-ph.CO]} \BibitemShut {NoStop}%
\end{thebibliography}%


%
%


\end{document}